\documentclass[onecolumn,a4paper,12pt,oneside]{article} 
\usepackage[top=2.5cm,left=2.5cm,right=2.5cm,bottom=3cm]{geometry}    
\usepackage{draftwatermark} 
\SetWatermarkText{\bf PREPRINT}
\SetWatermarkAngle{45}
\SetWatermarkScale{2.5}
\SetWatermarkLightness{0.85}     
\usepackage[parfill]{parskip}    
\usepackage{graphicx}
\usepackage{amssymb}
\usepackage{amsmath}
\usepackage{amsfonts}
\usepackage{amsthm}
\usepackage{epstopdf}
\usepackage[usenames,dvipsnames,svgnames,table]{xcolor}
\usepackage{tikz}
\usepackage{array}
\usepackage[T1]{fontenc}
\usepackage[utf8]{inputenc}
\usepackage{booktabs} 
\usepackage{array} 
\usepackage{paralist} 
\usepackage{listings} 
\usepackage{verbatim} 
\usepackage{subfig} 
\usepackage[colorlinks=true]{hyperref}
\hypersetup{colorlinks=true, linktocpage=true, linkcolor=Blue, citecolor=Green,
hypertexnames=true, urlcolor=Blue,pdftex}
\usepackage[english]{babel}
\usepackage{wrapfig}
\usepackage{multirow}
\usepackage{quoting}
\usepackage{rotating}
\quotingsetup{font=normalsize}
\theoremstyle{plain}

\usepackage{bm}
\usepackage{abstract}
\usepackage{cite}
\usepackage{float}
\usepackage{textcomp} 	
\usepackage{tabularx} 
\usepackage{caption}
\usepackage{authblk} 
\usepackage{eso-pic} 
\usepackage{enumitem} 
\usepackage{lineno} 
\providecommand{\keywords}[1]{\textbf{{Key words: }} #1} 

\usepackage{graphicx}
\usepackage{multirow}
\usepackage{amsmath,amssymb,amsfonts}
\usepackage{mathrsfs}
\usepackage[title]{appendix}
\usepackage{xcolor}
\usepackage{textcomp}
\usepackage{manyfoot}
\usepackage{booktabs}
\usepackage{algorithm}
\usepackage{algorithmicx}
\usepackage{algpseudocode}
\usepackage{listings}
\usepackage{paralist}

\newcommand{\be}{\begin{equation}}
\newcommand{\ee}{\end{equation}}
\newcommand{\bs}{\begin{split}}
\newcommand{\es}{\end{split}}

\renewcommand{\Phi}{\varPhi}

\newcommand{\Sy}{\mathbb{S}}

\renewcommand{\Theta}{\varTheta}
\renewcommand{\Psi}{\varPsi}
\renewcommand{\Sigma}{\varSigma}
\newcommand{\A}{\mathbb{A}}

\newcommand{\B}{\mathbb{B}}

\newcommand{\D}{\mathbb{D}}

\newcommand{\Q}{\mathbb{Q}}

\renewcommand{\Delta}{\varDelta}
\renewcommand{\phi}{\varphi}
\renewcommand{\psi}{\varPsi}

\newcommand{\nud}{\nu_{12}}

\title{\LARGE{\bf Anisotropic auxetic composite laminates: a polar approach}}

\author{Paolo Vannucci\smallskip\\
\begin{small}{ LMV - Laboratoire de Mathématiques de Versailles, UMR8100 \\
 Université de Versailles et Saint Quentin - 45, Avenue des Etats-Unis, 78035 - France\\
           \href{mailto:paolo.vannucci@uvsq.fr}{paolo.vannucci@uvsq.fr}}  \bigskip\bigskip\\
           Final version in Journal of Composite Materials\\ \href{https://doi.org/10.1177/00219983241256335}{https://doi.org/10.1177/00219983241256335}\end{small}}

\begin{document}

\maketitle

\begin{abstract}
{The problem of obtaining anisotropic auxetic composite laminates, i.e. having a negative Poisson's ratio for at least some directions, is examined in this paper. In particular, the possibility of obtaining auxeticity  stacking uni-directional identical plies is considered. It is shown that if the ply is composed by isotropic matrix and fibers, then it is impossible to obtain totally auxetic orthotropic laminates, i.e. auxeticity for each direction, unless at least one among matrix and fibers is auxetic itself. Moreover, it is shown what are the conditions, in terms of the mechanical properties of the constituents and of the volume fraction of the fibers,  to fabricate uni-directional plies with which to realize laminates having a negative Poisson's ratio for some directions.  Several existing materials are also examined. 
All the analysis is done using the polar formalism, very effective for the study of plane anisotropic problems.}
\end{abstract}

\keywords{Poisson's ratio, auxeticity, anisotropy, composite laminates, polar formalism}

\section{Introduction}
The property of having a negative Poisson's ratio is known in the literature as {\it auxeticity}\cite{cho2019}. Theoretically physically possible  for isotropic materials, auxeticity is normally obtained if some peculiar material microstructure is present in an isotropic (at least at a sufficiently large scale)  continuum. Several studies have been done the last three decades on the possibility of obtaining auxetic materials by the use of some peculiar microstructures, also at the molecular level, from the pioneer works of Almgren\cite{Almgren85}, Evans\cite{Evans91,Evans_Nature91}, Lakes\cite{Lakes1991,Lakes1993,Lakes2002,Lakes2017}, Milton\cite{Milton92}. Literature on the topic is very huge, it is not  the objective here to give a complete overview of the matter and the reader is addressed to specific review articles on auxetic materials, like the papers of Prawoto\cite{Prawoto12} or of  Shukla \& Behera\cite{Shukla22}. 

\begin{figure}
\centering
\includegraphics[width=.4\columnwidth]{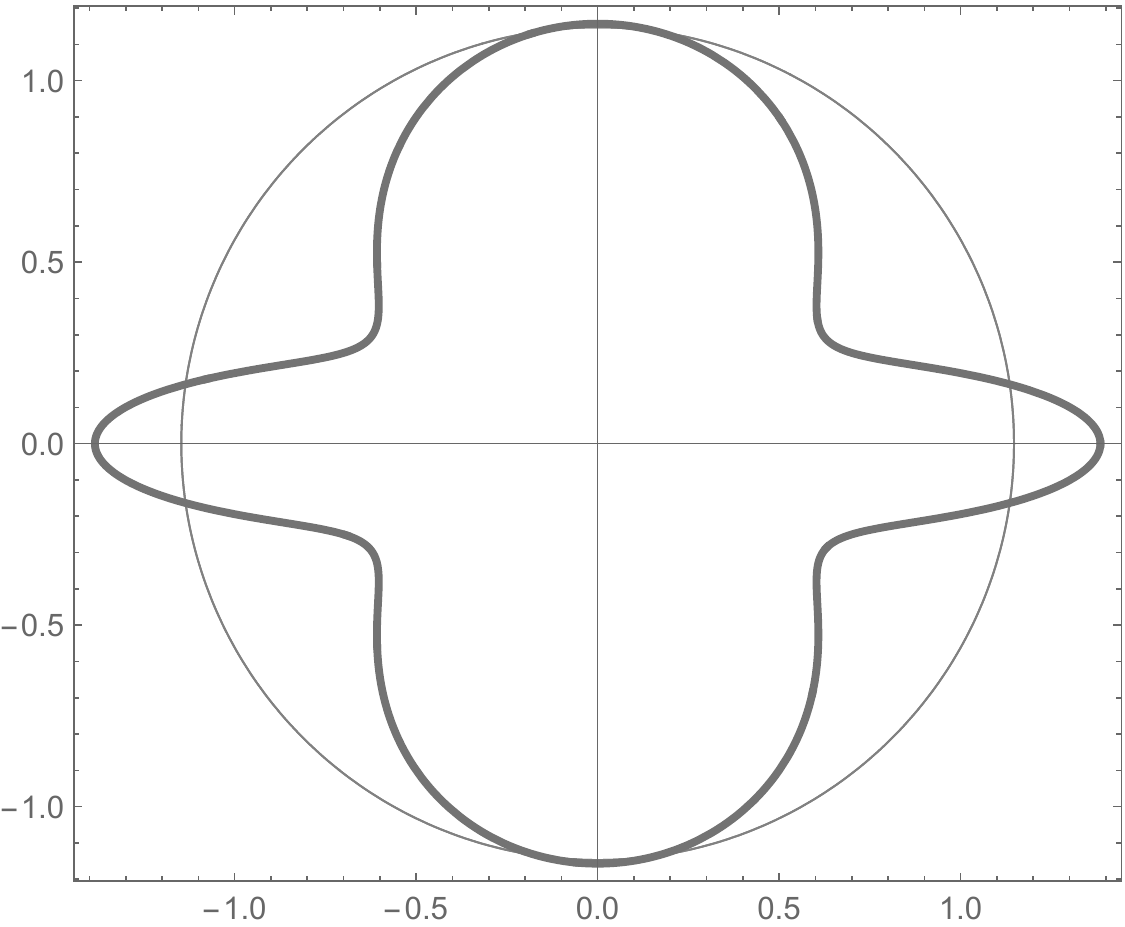}
\caption{The directional diagram of $\nu_{12}$ for pine wood in a plane-stress state (material data in Tab. \ref{tab:1}, source Lekhnitskii\cite{Lekhnitskii}). The thin  circle marks the zero value: inside it, $\nud$ is negative.}
\label{fig:1}
\end{figure}

For an anisotropic body, also auxeticity is an anisotropic property, i.e. the Poisson's coefficient can be negative for some directions, positive for some others, and this without the need of  any peculiar microstructure:  the directional variations typical of anisotropy make the auxeticity of classical elastic  materials  {\it à la Cauchy} possible. For example, the Poisson's ratio $\nu_{12}$ for a board of pine wood in a plane-stress state is  negative  for almost all the directions, Fig. \ref{fig:1}.
Because anisotropic composite materials are mainly used in the fabrication of laminated structures, it has hence a certain interest the possibility of obtaining auxetic laminates. A recent state of the art on auxetic composite laminates for structural uses has been published  by Veloso et al\cite{Veloso23}. The first studies on auxetic anisotropic laminates date back to the eighties\cite{Herakovich84,Miki89}. Clarke at al\cite{Clarke94} and Hine et al\cite{Hine97} investigated experimentally the auxeticity of balanced angle-ply laminates, a type of laminates frequently considered for the simplicity of the stack. The possibility of obtaining auxetic laminates is considered by Zhang et al\cite{Zhang98,Zhang99} and by Alderson et al\cite{alderson2005}, while the determination of the highest negative  Poisson's ratio is studied in\cite{Peel07,Shokrieh11}.

However, in all the existing literature on composite laminates, the problem of the auxeticity of the elastic response is treated only partially, with regard to some specific cases, i.e. angle-ply stacks, or concerning some special situations or aspects, like for instance the constituent materials or the maximization of the negative value of the Poisson's ratio. The aim of this paper is precisely to give some generality to the matter and namely to try to respond as largely as possible to the following fundamental question: {\it when is it possible to fabricate an auxetic laminate stacking identical anisotropic plies}? In other terms, given an anisotropic ply, will it exist at least one stacking sequence giving rise to a laminate that will have an auxetic response? This leads to investigate the relations existing between the {\it material properties} of the ply and the {\it geometry} of the stack. We want to stress that we look for an auxetic behavior {\it generated uniquely by anisotropy}, i.e. the component layers as well as their constituent materials are classical elastic bodies {\it à la Cauchy}, with no  microstructures able to generate auxeticity, so just the classical anisotropic elasticity theory  is used for this study.

The study proposed here concerns  only the in-plane properties of the laminate, not  the transversal ones. In addition, we consider laminates composed by identical unidirectional layers (the case of plies reinforced by fabrics will be the subject of a future study). Moreover, only orthotropic laminates are considered, as  this is the case for usual structures. The results are presented for the extension behavior, but actually they can be rephrased {\it verbatim} for the bending one (the auxeticity of the bending response is less interesting for applications). 

The problem is rather complicate and it reveals interesting, mathematically speaking, to formulate it in an appropriate space of material and geometrical parameters. In particular, for representing the material properties this study makes use of  dimensionless  moduli deduced from the {\it polar formalism} for plane anisotropy, introduced by G. Verchery in 1979\cite{Verchery79,Vannucci05,vannucci_libro} and of  the, also dimensionless, {\it lamination parameters} proposed in 1968 by Tsai and Pagano\cite{tsai1968} for the geometry of the stack.

In the paper, four different theoretically possible cases of in-plane auxeticity for anisotropic laminates are introduced, all of them are considered and analyzed. Several existing materials are also analyzed and the question whether or not a type of auxeticity can be obtained for the laminate using classical anisotropic plies, i.e. plies made of non auxetic materials, is considered. Some examples of auxetic laminates are also given. 

\section{Fundamental relations}
\subsection{The in-plane Poisson's ratio $\nud$} 
Let us consider an anisotropic layer in a  plane-stress state, with $\Sy$ and $\Q=\Sy^{-1}$ respectively the compliance and the reduced stiffness  tensors. The in-plane Poisson's ratio $\nud(\theta)$ at the direction inclined of $\theta$ on the $x_1-$axis is defined as\cite{Lekhnitskii,TsaiHahn,Jones,Ting} (we adopt in this study the Kelvin's notation\cite{kelvin,kelvin1} for the elasticity tensors)
\be
\nud(\theta):=-\frac{\Sy_{12}(\theta)}{\Sy_{11}(\theta)}.
\ee
Because $\Sy_{11}(\theta)>0\ \forall\theta$, 
\be
\nud(\theta)<0\iff\Sy_{12}(\theta)>0.
\ee
\subsection{Polar formulation of the auxeticity condition}
In the polar formalism, the above condition for an orthotropic layer becomes\cite{vannucci_libro}
\be
\label{eq:auxeticity0}
\Sy_{12}(\theta)=-t_0+2t_1- r_0\cos4(\phi_0-\theta)>0,
\ee
with $t_0, t_1,r_0$ non-negative tensor invariants depending upon the components of $\Sy$ and $\phi_0$ a polar angle. In view of the forthcoming developments, it is worth to transform the previous relation. To this end, we express  the polar parameters of the compliance $\Sy$ by those of the stiffness $\Q$:
\be
\begin{array}{l}
t_0=2\dfrac{T_0T_1-R_1^2}{\Delta},\medskip\\
t_1=\dfrac{T_0^2-R_0^2}{2\Delta},\medskip\\
r_0 \mathrm{e}^{4\mathrm{i}\phi_0}=\dfrac{2}{\Delta}(R_1^2\mathrm{e}^{4\mathrm{i}\Phi_1}-T_1R_0\mathrm{e}^{4\mathrm{i}\Phi_0}),
\end{array}
\ee
where $T_0,T_1,R_0,R_1$ are the polar invariant (positive) moduli of $\Q$ and $\Phi_0,\Phi_1$ the two polar angles of $\Q$, whose difference is the fifth invariant. $\Delta$ is the invariant given by
\be
\Delta=4T_1(T_0^2-R_0^2)-8R_1^2[T_0-R_0\cos4(\Phi_0-\Phi_1)].
\ee
Some simple passages give
\be
\begin{split}
r_0\cos4(\phi_0-\theta)&=\frac{2}{\Delta}[R_1^2\cos4(\Phi_1-\theta)-\\
&-T_1R_0\cos4(\Phi_0-\theta)].
\end{split}
\ee
As, actually, $\Delta=\det\Q>0$, it can be ignored in the following, because it does not change the sign of $\Sy_{12}$. So, in terms of the polar parameters of $\Q$ the auxeticity condition is
\be
\bs
\hspace{-2mm}2(T_0T_1-R_1^2)-T_0^2+R_0^2+2[R_1^2\cos4(\Phi_1-\theta)-&\\-T_1R_0\cos4(\Phi_0-\theta)]&<0.
\end{split}
\ee
A rotation $\alpha$ of the frame corresponds to subtract $\alpha$ from the two polar angles. If $\alpha=\Phi_1$ or, equivalently, if the reference frame is chosen in such a way that $\Phi_1=0$, which corresponds, for unidirectional (UD) plies, to put the $x_1-$axis aligned with the fibres, the above relation becomes
\be
\bs
2(T_0T_1-R_1^2)-T_0^2+R_0^2+2[R_1^2\cos4\theta-&\\-T_1R_0\cos4(\Phi-\theta)]&<0,
\end{split}
\ee
with $\Phi=\Phi_0-\Phi_1$, the angular invariant of $\Q$. Common orthotropy corresponds to the condition\cite{vannucci_libro} 
\be
\Phi=K\frac{\pi}{4},\ \ K\in\{0,1\},
\ee
the value of $K$ determining two different types of orthotropic materials sharing the same polar moduli of $\Q$. As said in the Introduction, we  consider in this paper only UD plies (i.e. layers reinforced by fabrics are not considered). It can be shown that such plies can be only  ordinary orthotropic hence $R_0\neq0,R_1\neq0$.

So finally, for an UD layer, the auxeticity condition resumes to
\be
\label{eq:auxUD}
\bs
\lambda(\theta):&=2(T_0T_1-R_1^2)-T_0^2+R_0^2+\\&+2[R_1^2-(-1)^KT_1R_0]\cos4\theta<0,
\end{split}
\ee

\subsection{Auxeticity condition for an orthotropic laminate}
Let us now consider a laminate composed of identical UD plies all sharing the same reduced stiffness tensor $\Q$. We focus on the extension response, described by tensor\cite{Jones,vannucci_libro}
\begin{equation}
\label{eq:Atens}
\A=\frac{1}{h}\sum_{j=1}^n({z_j}-{z_{j-1}})\Q_(\delta_j),\\
\end{equation}
with $\delta_j$ the orientation of the $j-$th layer among the $n$ composing the laminate and $h$ the plate's thickness. 
We further assume that the laminate is extension-bending uncoupled. This assumption is necessary, because otherwise the Poisson's ratio for the extension, or also for the bending, behavior, should be practically impossible to be analyzed (in fact, for coupled laminates the compliances, in extension and in bending, depend in a very complicate manner upon $\A,\B$ and $\D$, respectively the stiffness tensors in extension, coupling and bending\cite{vannucci01joe,vannucci23a,vannucci23b}). It is worth to recall that uncoupling ($\B=\mathbb{O}$) can be obtained by suitable stacking sequences, but, contrarily to what commonly believed, not necessarily symmetric. Actually, asymmetric uncoupled laminates are much more numerous than the symmetric ones\cite{vannucci01ijss,vannucci01cst}. So, uncoupling can be obtained rather easily and to assume it does not constitute a true limitation.

The auxeticity condition for an orthotropic tensor $\A$  is the same of that for $\Q$, provided that the polar parameters of $\A$ (denoted in the following  by a superscript $A$) replace those of the layer:
\be
\label{eq:lambdaA0}
\bs
\lambda^A(\theta):&=2(T_0^AT_1^A-{R_1^A}^2)-{T_0^A}^2+{R_0^A}^2+\\&+2[{R_1^A}^2-(-1)^{K^A}T_1^AR_0^A]\cos4\theta<0.
\end{split}
\ee

Because the plies are identical, the polar parameters of the extension stiffness tensor $\A$  can be put in the form\cite{vannucci_libro}
\begin{equation}
\label{eq:compApolareq}
\begin{array}{l}
T_0^A=T_0,\medskip\\
T_1^A=T_1,\medskip\\
R_0^A\mathrm{e}^{4\mathrm{i}\varPhi_0^A}={R_0\mathrm{e}^{4\mathrm{i}\varPhi_0}}{\left(\xi_1+\mathrm{i}\xi_2\right)},\medskip\\
R_1^A\mathrm{e}^{2\mathrm{i}\varPhi_1^A}={R_1\mathrm{e}^{2\mathrm{i}\varPhi_1}}{\left(\xi_3+\mathrm{i}\xi_4\right)}.
\end{array}
\end{equation}
The quantities $\xi_i,i=1,...,4$ are the {\it lamination parameters}\cite{tsai1968} for $\A$:
\begin{equation}
\label{eq:laminationparameters}
\xi_1+\mathrm{i}\xi_2=\frac{1}{n}\sum_{j=1}^n\mathrm{e}^{4\mathrm{i}\delta_j},\ \ \xi_3+\mathrm{i}\xi_4=\frac{1}{n}\sum_{j=1}^n\mathrm{e}^{2\mathrm{i}\delta_j},
\end{equation}
Eq. (\ref{eq:compApolareq}) shows, on the one hand, that the isotropic part of $\A$ is identical to that of the layer, $T_0$ and $T_1$,  and, on the other hand, because the layer is UD, if we consider, as said above, an orthotropic extension behavior, choosing $\Phi^A_1=0$ to fix the reference frame for the laminate, we get easily
\be
\begin{array}{c}
\xi_2=\xi_4=0,\medskip\\ (-1)^{K^A}R_0^A=(-1)^KR_0\xi_1,\medskip\\ R_1^A=R_1\xi_3.
\end{array}
\ee
So, the polar auxeticity condition for $\A$ becomes
\be
\label{eq:lambdaA1}
\bs
\lambda^A(\theta)&=2(T_0T_1-R_1^2\xi_3^2)-T_0^2+R_0^2\xi_1^2+\\&+2[R_1^2\xi_3^2-(-1)^KT_1R_0\xi_1]\cos4\theta<0.
\end{split}
\ee

\subsection{Dimensionless auxeticity condition}
In the expression of $\lambda^A(\theta)$, there are three kind of variables/parameters: the independent variable $\theta$, determining the {\it direction}, the two lamination parameters $\xi_1$ and $\xi_3$, accounting for the {\it geometry}  of the stack (i.e., the sequence of the orientation angles $\delta_j$), and the polar parameters $T_0,T_1,(-1)^KR_0$ and $R_1$ of the layer, representing the {\it material} part of $\lambda^A(\theta)$. It is worth to introduce new parameters for the material part, dimensionless like $\xi_1$ and $\xi_3$. This will reduce the number of independent material parameters and allow an easier study of the relation between geometry and material. To this end, we introduce the following ratios\cite{vannucci13}:
\be
\label{eq:adimpar}
\tau_0=\frac{T_0}{R_1},\ \ \tau_1=\frac{T_1}{R_1},\ \ \rho=\frac{R_0}{R_1}.
\ee
It is worth noting that these ratios can be introduced because, for a UD ply, $R_1\neq0$. 
The auxeticity condition (\ref{eq:lambdaA1}) can be rewritten in the equivalent form
\be
\label{eq:lambdaA2}
\bs
\lambda^A(\theta)&=2(\tau_0\tau_1-\xi_3^2)-\tau_0^2+\rho^2\xi_1^2+\\&+2[\xi_3^2-(-1)^K\tau_1\rho\ \xi_1]\cos4\theta<0,
\end{split}
\ee
depending on five dimensionless quantities: three for the ply, material part, and two for the stack, geometric part. 

We present in Tab. \ref{tab:1} the characteristics of twenty different UD plies commonly used for the fabrication of composite laminates.

\begin{table}
\centering\footnotesize\sf
\caption{Some examples of UD plies. Modules are in GPa, $\Phi_0=\Phi_1=0\Rightarrow K=0$ for all the plies.}
\begin{tabular}{rrrrrrrrrrrr}
\toprule
Mat.&$E_1$&$E_2$&$G_{12}$&$\nu_{12}$&$T_0$&$T_1$&$R_0$&$R_1$&$\tau_0$&$\tau_1$&$\rho$\\
\midrule
1&10.00&0.42&0.75&0.24&1.66&1.34&0.91&1.20&1.383&1.116&0.758\\
2&181.00&10.30&7.17&0.28&26.88&24.74&19.71&21.43&1.254&1.154&0.919\\
3& 205.00&18.50&5.59&0.23&29.80&29.14&24.21&23.42&1.272&1.244&1.033\\
4&38.6&8.27&4.14&0.26&7.47&6.49&3.33&3.85&1.940&1.686&0.865\\
5&86.90&5.52&2.14&0.34&12.23&12.11&10.09&10.25&1.193&1.181&0.984\\
6&47.66&13.31&4.75&0.27&9.24&8.70&4.49&4.38&2.108&1.984&1.023\\
7&54.00&18.00&9.00&0.25&12.54&10.34&3.54&4.59&2.729&2.250&0.771\\
8&207.00&5.00&2.60&0.25&27.52&26.85&24.93&25.29&1.088&1.062&0.986\\
9&76.00&5.50&2.10&0.34&10.85&10.74&8.75&8.88&1.221&1.209&0.984\\
10&207.00&21.00&7.00&0.30&30.67&30.35&23.67&23.46&1.307&1.293&1.009\\
11&45.00&12.00&4.50&0.30&8.62&8.22&4.13&4.22&2.041&1.945&0.976\\
12&134.00&7.00&4.20&0.25&19.34&18.12&15.14&15.93&1.214&1.138&0.951\\
13&85.00&5.60&2.10&0.34&11.98&11.89&9.88&10.00&1.198&1.189&0.988\\
14&220.00&140.00&7.50&0.25&41.50&55.98&34.00&10.41&3.985&5.375&3.265\\
15&294.50&6.34&4.90&0.23&39.73&38.01&34.83&36.06&1.102&1.054&0.966\\
16&109.70&8.55&5.31&0.30&16.89&15.53&11.58&12.73&1.327&1.220&0.910\\
17&131.70&8.76&5.03&0.28&19.55&18.26&14.52&15.45&1.265&1.182&0.940\\
18&133.10&9.31&3.74&0.34&19.02&18.74&15.28&15.60&1.219&1.201&0.979\\
19&135.00&9.24&6.28&0.32&20.55&18.89&14.27&15.83&1.298&1.193&0.901\\
20&128.00&13.00&6.40&0.30&20.00&18.77&13.60&14.51&1.378&1.293&0.937\\
\bottomrule
\end{tabular}\\
\begin{footnotesize}
\begin{tabular}{ll}
1: Pine wood\cite{Lekhnitskii}&11: Glass-epoxy\cite{gay14}\\
2: Carbon-epoxy T300/5208\cite{TsaiHahn}&12: Carbon-epoxy\cite{gay14}\\
3: Boron-epoxy B(4)-55054\cite{TsaiHahn}&13: Kevlar-epoxy\cite{gay14}\\
4: Glass-epoxy s-ply1002 V$_\mathrm{f}$=0.45\cite{TsaiHahn}&14: Boron-aluminium\cite{gay14}\\
5: Kevlar-epoxy 149\cite{daniel94}&15: Carbon-epoxy GY70/34\cite{MILHDBK}\\
6: S-glass-epoxy S2-449/SP381\cite{MILHDBK}&16: Carbon-bismaleide AS4/5250-3\cite{MILHDBK}\\
7: Glass-epoxy\cite{Jones}&17: Carbon-peek AS4/APC2\cite{MILHDBK}\\
8: Carbon-epoxy\cite{Jones}&18: Carbon-epoxy AS4/3502\cite{MILHDBK}\\
9: Kevlar-epoxy\cite{Jones}&19: Carbon-epoxy T300/976\cite{MILHDBK}\\
10: Boron-epoxy\cite{Jones}&20: Carbon-epoxy 3 MXP251S\cite{Gurdal99}\\
\end{tabular}
\end{footnotesize}
\label{tab:1}
\end{table}

\subsection{Thermodynamic necessary conditions}
The polar moduli, like any other set of parameters representing elasticity, must satisfy some thermodynamic conditions, stating the positivity of the elastic potential\cite{Love,Ting,vannucci_libro,vannucci24}. In\cite{vannucci13} it has been shown that using the dimensionless parameters (\ref{eq:adimpar}), for an orthotropic ply these conditions are
\be
\label{eq:thermo}
\bs
&\tau_0-\rho>0,\\
&\tau_1\left[\tau_0+(-1)^K\rho\right]-2>0.
\end{split}
\ee
They are necessary for establishing the set, in the space of the polar parameters, of materials giving rise to auxetic laminates, see below.

\subsection{Theoretical considerations}
An interesting interpretation can be given to the auxeticity condition: Eq. (\ref{eq:auxeticity0}) can equivalently be rewritten as
\be
\frac{1}{4[t_0+r_0\cos4(\phi_0-\theta)]}>\frac{1}{8t_1}.
\ee
The two members of the above inequality have a direct mechanical interpretation\cite{vannucci_libro}: the shear modulus $G_{12}(\theta)$ is
\be
\bs
G_{12}(\theta)&=\frac{1}{4[t_0-r_0\cos4(\phi_0-\theta)]}=\\
&=\frac{1}{4[t_0+r_0\cos4(\phi_0-\theta-\frac{\pi}{4})]}.
\end{split}
\ee
The bulk modulus $\kappa$ is
\be
\kappa=\frac{1}{8t_1}.
\ee
So, the auxeticity condition for the direction $\theta+\frac{\pi}{4}$ actually corresponds to state that
\be
G_{12}(\theta)>\kappa.
\ee
This condition also includes and generalizes the already discussed auxeticity condition\cite{Milton92,Prawoto12} for isotropy ($r_0=0$), that in 2D elasticity reads like:
\be
\frac{1}{4t_0}>\frac{1}{8t_1}\Rightarrow G>\kappa.
\ee

\subsection{The lamination domain}
\begin{figure}
\centering
\includegraphics[width=.5\columnwidth]{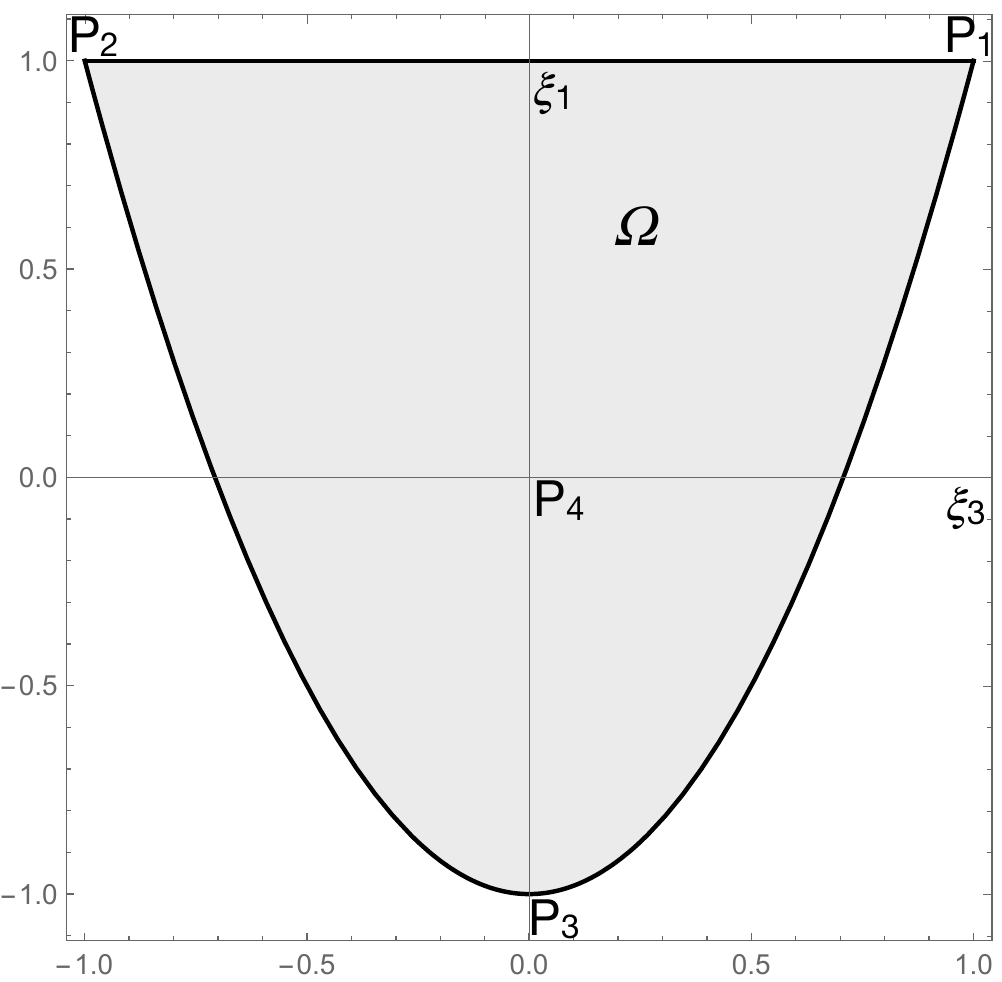}
\caption{The lamination domain of an orthotropic $\A$.}
\label{fig:2}
\end{figure}

Miki\cite{Miki82,Miki83,Jones} has shown that  the {\it lamination domain} of an orthotropic $\A$, i.e. the set of points of the plane $(\xi_3,\xi_1)$, that can correspond to an orthotropic behavior in extension, is the sector of parabola $\Omega$ in Fig. \ref{fig:2} defined by the bounds
\be
2\xi_3^2-1\leq\xi_1\leq1,\ \ -1\leq\xi_1\leq1.
\ee
Each lamination point $(\xi_3,\xi_1)$ corresponds to an extension response, i.e. to a tensor $\A$, that, in general, can be realized by more that one stacking sequence.
We just recall that the lamination point P$_1=(1,1)$ corresponds to a unidirectional laminate with all the plies with $\delta_j=0\ \forall j$, P$_2=(-1,1)$ to a unidirectional laminate with $\delta_j=\frac{\pi}{2}\forall j$, P$_3=(0,-1)$ to a balanced angle ply with $\delta_j=\pm\frac{\pi}{4}$, P$_4=(0,0)$ to an isotropic laminate. Moreover, any angle-ply laminate is represented by a point of the parabolic boundary  and any  cross-ply laminate by a point of the line P$_1-$P$_2$. 

Miki\cite{Miki85} has also shown that a tensor $\D$ describing the bending behavior has exactly the same lamination domain, although the definition of the relevant lamination parameters is different. That is why all the results of this paper, concerning extension, can be exported identically to bending, the stacking sequence apart. 

\section{Auxeticity conditions for an orthotropic laminate}
As mentioned above, for anisotropic bodies also auxeticity is an anisotropic property. However, when an UD ply is used to fabricate a laminate, it is of interest, at least theoretically, to examine the different situations that can determine an auxetic behavior of the laminate. First of all, a natural question is: because physics allows auxetic isotropic materials ($-1<\nu<0$), one can wonder whether or not it is possible to realize, stacking identical UD plies, a Totally Auxetic Anisotropic Laminate (TAAL), i.e. an anisotropic, or also isotropic,  laminate that is auxetic for each direction. Secondarily, one could consider a relaxed requirement: an anisotropic laminate that should be auxetic for some directions, situation that in short we will denote by the acronym PAAL (Partially Auxetic Anisotropic Laminate). For a given UD material, four different situations are, in principle, possible:
\begin{enumerate}
\item the material allows to fabricate a TAAL for each possible lamination point $(\xi_3,\xi_1)\in\Omega$;
\item the material allows to fabricate a TAAL for some lamination points $(\xi_3,\xi_1)\in\Omega$;
\item the material allows to fabricate a PAAL for each possible lamination point $(\xi_3,\xi_1)\in\Omega$;
\item the material allows to fabricate a PAAL for some lamination points $(\xi_3,\xi_1)\in\Omega$.
\end{enumerate}
The objective of this paper is precisely to determine the conditions to be satisfied by a material in order to obtain one of the four cases above. It is a matter of fact that composite anisotropic auxetic laminates can exist\cite{Veloso23}, but the general conditions allowing to obtain them and, more precisely, of what type among the four cases above, has never been elucidated in the literature, only some partial recommandations have been proposed. We aim at giving a general response through the use of classical anisotropic elasticity and the polar formalism.

 \subsection{The conditions for obtaining a TAAL}
\begin{figure}
\centering
\includegraphics[width=\columnwidth]{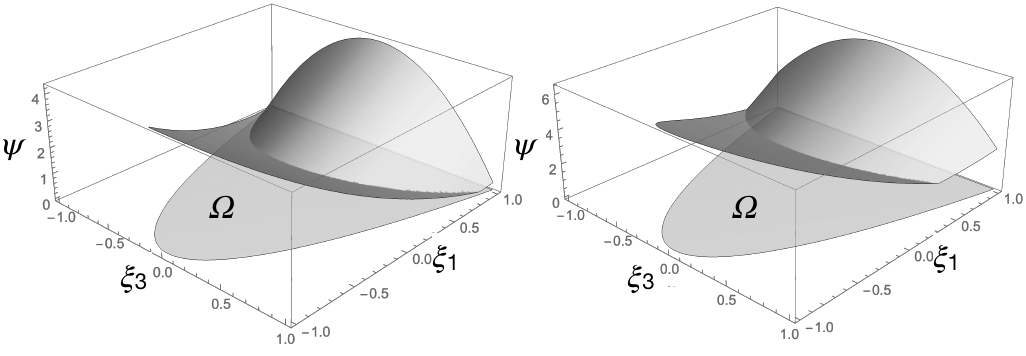}
\caption{The function $\psi(\xi_3,\xi_1)$ on $\Omega$ for materials 2 and 4 of Tab. \ref{tab:1}.}
\label{fig:3}
\end{figure}

\begin{figure}
\centering
\includegraphics[width=.5\columnwidth]{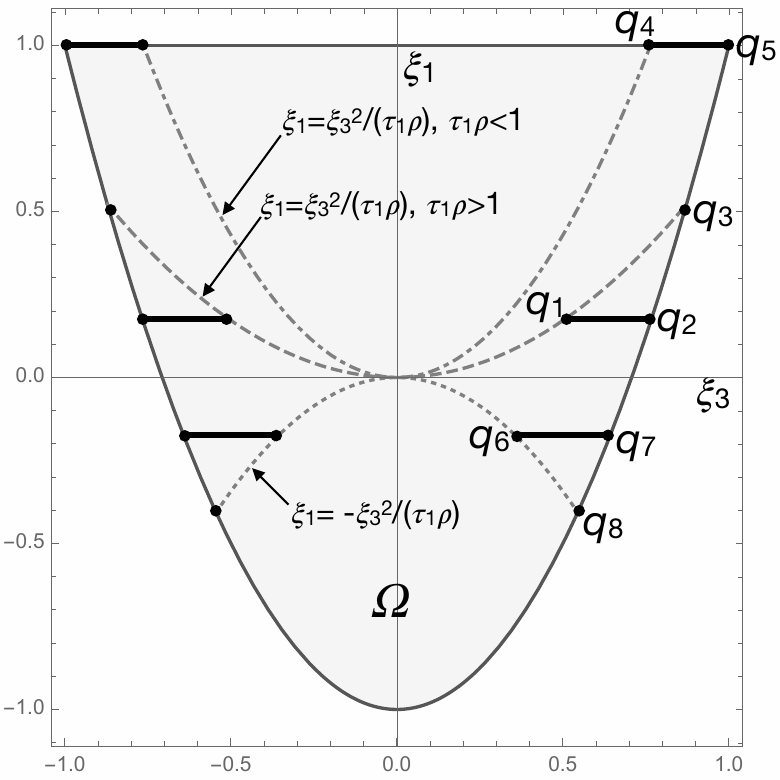}
\caption{The points of minimum of the function $\psi(\xi_3,\xi_1)$.}
\label{fig:4}
\end{figure}

\begin{figure}
\centering
\includegraphics[width=.6\columnwidth]{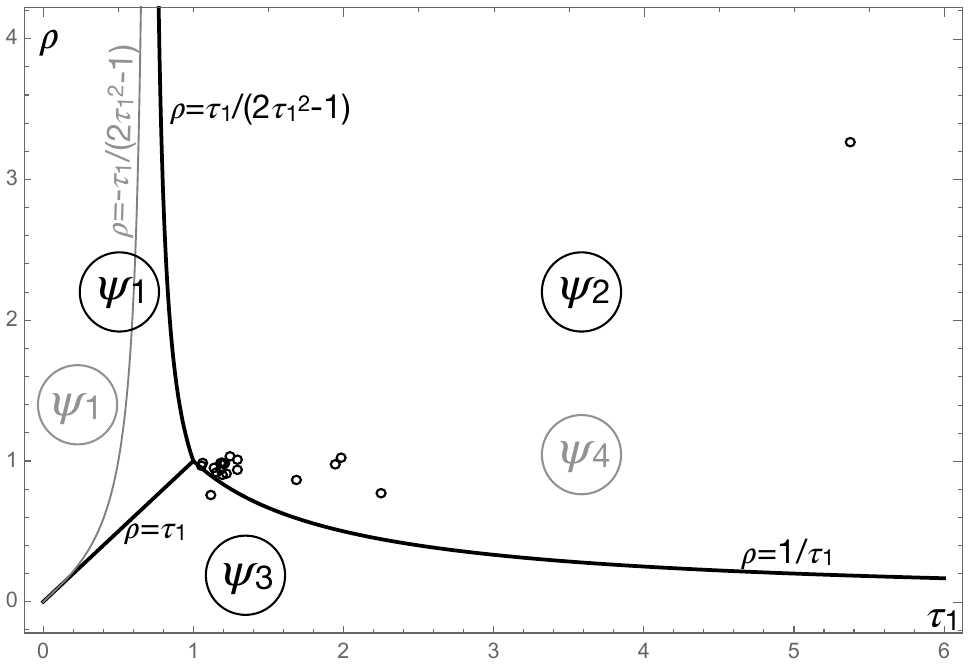}
\caption{The zones of the minimum of $\psi(\xi_3,\xi_1)$ for the case $K=0$, in black, and $K=1$, in grey. Each circle corresponds to a material of Tab. \ref{tab:1}.}
\label{fig:5}
\end{figure}

 For an anisotropic laminate to be totally auxetic, equation (\ref{eq:lambdaA2}) must be satisfied $\forall\theta\in\left[0,\frac{\pi}{2}\right]$. But $\lambda^A(\theta)\leq0\ \forall\theta\iff\max\lambda^A(\theta)\leq0$ which gives the condition
\be
\label{eq:psi}
\bs
\psi(\xi_3,\xi_1):&=2(\tau_0\tau_1-\xi_3^2)-\tau_0^2+\rho^2\xi_1^2+\\&+2|\xi_3^2\hspace{-1mm}-\hspace{-1mm}(-1)^K\tau_1\rho\ \xi_1|\hspace{-1mm}<0.
\end{split}
\ee
For a chosen ply, the function $\psi(\xi_3,\xi_1)$ is defined on $\Omega$; it is shown in Fig. \ref{fig:3} for materials 2 and 4 in Tab. \ref{tab:1}.
Then:
\begin{itemize}
\item an UD ply can fabricate a TAAL $\forall(\xi_3,\xi_1)\in\Omega\iff\max\ \psi(\xi_3,\xi_1)<0$;
\item an UD ply can produce a TAAL for some lamination points $(\xi_3,\xi_1)\in\Omega\iff\min\ \psi(\xi_3,\xi_1)<0$.
\end{itemize}
The  maximum and minimum of $\psi(\xi_3,\xi_1)$ depend on the material parameters $\tau_0,\tau_1$ and $\rho$ and are rather cumbersome though not difficult to be found. The maximum has the same expression regardless the type of orthotropy, $K=0$ or $K=1$:
\be
\psi_{\max}=\psi(0,\pm1)=(\tau_0+\rho)(2\tau_1-\tau_0+\rho).
\ee
So, because the polar moduli are non-negative quantities, the condition for obtaining a TAAL $\forall(\xi_3,\xi_1)$ is
\be
\label{eq:TAAL1}
\psi_0:=2\tau_1-\tau_0+\rho<0.
\ee
The minimum can take different values and depends also on the type of orthotropy. If $K=0$,
\be
\begin{array}{l}
\psi_{\min}=-(\tau_0-\tau_1)^2:=\psi_1\ \ \ \ \mathrm{if }\ \ \tau_1<\rho<\dfrac{\tau_1}{2\tau_1^2-1},\medskip\\
\psi_{\min}=\left(\dfrac{\rho}{2\tau_1\rho-1}\right)^2-\dfrac{2\tau_1\rho}{2\tau_1\rho-1}-\tau_0^2+2\tau_0\tau_1:=\psi_2\medskip\\
\hspace{9mm}\mathrm{if }\ \ \ \rho>\dfrac{1}{\tau_1}\ \ \mathrm{and}\ \ \dfrac{\rho(2\tau_1^2-1)-\tau_1}{2\tau_1\rho-1}>0,\medskip\\
\psi_{\min}=(\tau_0-\rho)(2\tau_1-\tau_0-\rho):=\psi_3\medskip\\
\hspace{9mm} \mathrm{if }\ \ \ \rho<\tau_1\ \ \mathrm{and}\ \ \rho<\dfrac{1}{\tau_1}.
\end{array}
\ee
The condition for obtaining a TAAL for some points $(\xi_3,\xi_1)$ is hence, for $K=0$ orthotropic materials,
\be
\psi_{\min}\in\{\psi_1,\psi_2,\psi_3\}<0.
\ee
Moreover, referring to Fig. \ref{fig:4}: $\psi_1$ is get on the  segment $q_1q_2$ and on its symmetric with respect to the $\xi_1-$axis; $\psi_2$ is get on the point $q_3$ and on its symmetric; $\psi_3$ is get on the segment $q_4q_5$ and on its symmetric.
The coordinates of the points $q_i$ are:
\be
\begin{array}{l}
q_1=\left(\tau_1,\dfrac{\tau_1}{\rho}\right),\ q_2=\left(\sqrt{\dfrac{\rho+\tau_1}{2\rho}},\dfrac{\tau_1}{\rho}\right),\medskip\\ q_3=\left(\sqrt{\dfrac{\tau_1\rho}{2\tau_1\rho-1}},\dfrac{1}{2\tau_1\rho-1}\right),\medskip\\
q_4=\left(\sqrt{\tau_1\rho},1\right),\ q_5=(1,1).
\end{array}
\ee
A similar analysis can be done for materials with $K=1$:
\be
\begin{array}{l}
\psi_{\min}=\psi_1\ \ \ \ \mathrm{if }\  \ \ \ \rho(2\tau_1^2-1)+\tau_1<0,\medskip\\
\psi_{\min}=\left(\dfrac{\rho}{2\tau_1\rho+1}\right)^2-\dfrac{2\tau_1\rho}{2\tau_1\rho+1}-\tau_0^2+2\tau_0\tau_1:=\psi_4\medskip\\
\hspace{9mm}\mathrm{if }\ \ \  \rho(2\tau_1^2-1)+\tau_1>0,
\end{array}
\ee
so the condition for  obtaining a TAAL for some points $(\xi_3,\xi_1)$ using a $K=1$ orthotropic material is
\be
\psi_{\min}\in\{\psi_1,\psi_4\}<0.
\ee
Still referring to Fig. \ref{fig:4}, $\psi_1$ is get on the  segment $q_6q_7$ and on its symmetric with respect to the $\xi_1-$axis, while $\psi_4$ is get on the point $q_8$ and on its symmetric and the coordinates of these points  are:
\be
\begin{array}{l}
q_6=\left(\tau_1,-\dfrac{\tau_1}{\rho}\right),\ q_7=\left(\sqrt{\dfrac{\rho-\tau_1}{2\rho}},-\dfrac{\tau_1}{\rho}\right),\medskip\\ q_8=\left(\sqrt{\dfrac{\tau_1\rho}{2\tau_1\rho+1}},-\dfrac{1}{2\tau_1\rho+1}\right).
\end{array}
\ee
It is interesting to remark that, while $\psi_{\min}$ depends on all the material parameters, i.e., in the end, on the whole $\Q$, its position on $\Omega$ as well as the formula for its calculation depend only on $\tau_1$ and $\rho$.  The zones of the different $\psi_i,\ i=1,...,4,$ in the plane $(\tau_1,\rho)$ are depicted in Fig. \ref{fig:5}. On the same diagram, the positions of the materials in Tab. \ref{tab:1} are indicated by a small circle. It is apparent that all of them, of the type $K=0$, apart material  1, pine wood, belong to the zone where $\psi_{\min}=\psi_2$. Hence, for all of them, the minimum is get at the point $q_3$, on the boundary of $\Omega$. Such a lamination point can always be obtained by an angle-ply sequence, but not exclusively.

\subsection{The conditions for obtaining a PAAL}
 
\begin{figure}
\centering
\includegraphics[width=\columnwidth]{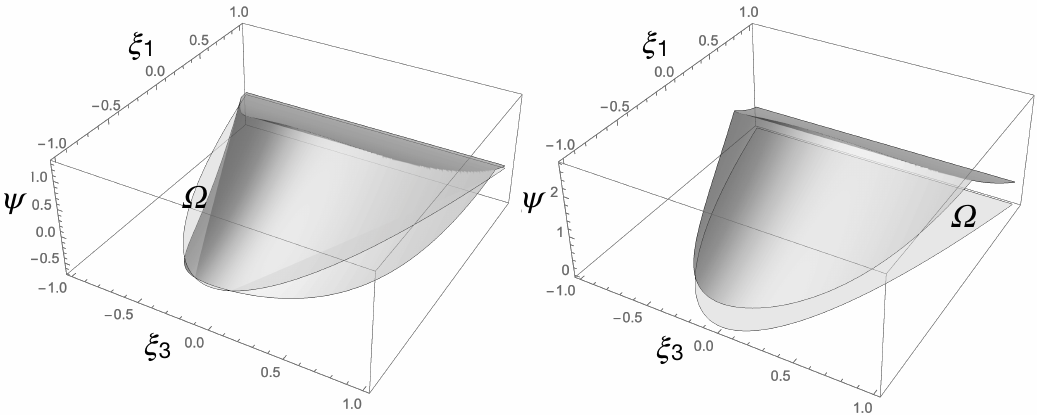}
\caption{The function $\eta(\xi_3,\xi_1)$ on $\Omega$ for materials 2 and 4 of Tab. \ref{tab:1}.}
\label{fig:6}
\end{figure}

\begin{figure}
\centering
\includegraphics[width=.6\columnwidth]{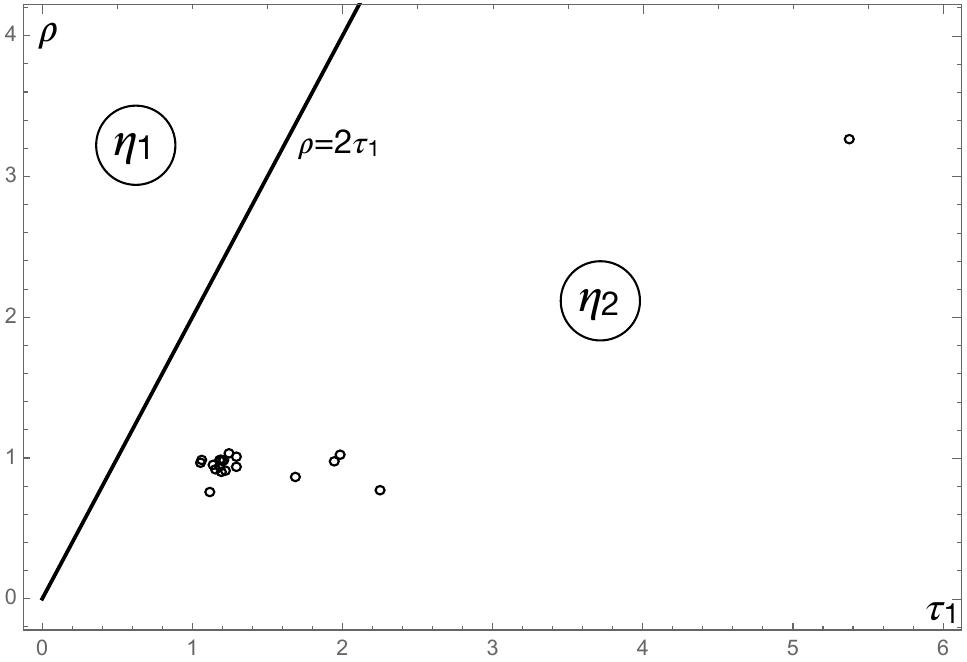}
\caption{The zones of the maximum of $\eta(\xi_3,\xi_1)$. Each circle corresponds to a material of Tab. \ref{tab:1}.}
\label{fig:7}
\end{figure}

An anisotropic laminate is partially auxetic if Eq. (\ref{eq:lambdaA2}) has at least one solution in $\Omega$. Hence, a PAAL can be fabricated with a UD material if and only if  $\min\lambda^A(\theta)\leq0$ which gives the condition
\be
\label{eq:psi}
\bs
\eta(\xi_3,\xi_1):&=2(\tau_0\tau_1-\xi_3^2)-\tau_0^2+\rho^2\xi_1^2-\\&-2|\xi_3^2\hspace{-1mm}-\hspace{-1mm}(-1)^K\tau_1\rho\ \xi_1|\hspace{-1mm}<0.
\end{split}
\ee
This case is rather similar to that of TAALs and can be treated in the same way. The
 function $\eta(\xi_3,\xi_1)$,  defined on $\Omega$, is shown in Fig. \ref{fig:6} for the same materials 2 and 4 in Tab. \ref{tab:1}.
It is evident that:
\begin{itemize}
\item an UD ply can fabricate a PAAL $\forall(\xi_3,\xi_1)\in\Omega\iff\max\ \eta(\xi_3,\xi_1)<0$;
\item an UD ply can produce a PAAL for some lamination points $(\xi_3,\xi_1)\in\Omega\iff\min\ \eta(\xi_3,\xi_1)<0$.
\end{itemize}
Like for function $\psi(\xi_3,\xi_1)$, the  maximum and minimum of $\eta(\xi_3,\xi_1)$ depend on the material parameters $\tau_0,\tau_1$ and $\rho$ and can be found by standard, though slightly articulated, differential calculus. 
Like for $\psi_{\max}$, also $\eta_{\max}$ is the same for $K=0$ and $K=1$ orthotropy:
\be
\begin{array}{l}
\eta_{\max}=(\tau_0-\rho)(2\tau_1-\tau_0-\rho):=\eta_1\ \mathrm{if}\ \rho\geq2\tau_1,\medskip\\
\eta_{\max}=\tau_0(2\tau_1-\tau_0):=\eta_2\ \mathrm{if}\ \rho<2\tau_1.
\end{array}
\ee
To remark that $\eta_1=\psi_3$; $\eta_1$ is get at $(0,-1)$ if $K=0$, at $(0,1)$ if $K=1$ i.e., in the two cases, for cross-ply balanced laminates, while $\eta_2$ is get at $(0,0)$ in both the cases, i.e. for isotropic laminates. The zones of  $\eta_1,\eta_2$ in the plane $(\tau_1,\rho)$ are shown in Fig. \ref{fig:7}, where   the positions of the materials in Tab. \ref{tab:1} are again denoted by a small circle.
The condition for obtaining a PAAL $\forall(\xi_3,\xi_1)\in\Omega$ is hence
\be
\eta_{\max}\in\{\eta_1,\eta_2\}<0.
\ee
Also   $\eta_{\min}$ can take different values and depends on the type of orthotropy. For the case  $K=0$,
\be
\label{eq:etamin1}
\begin{array}{l}
\eta_{\min}=\eta_1\ \ \mathrm{if}\ \ \rho^2-\tau_1\rho+1<0,\medskip\\
\eta_{\min}=(\tau_0+\rho)(2\tau_1-\tau_0+\rho)-4:=\eta_3\\
\hspace{29mm} \mathrm{if}\ \ \rho^2+\tau_1\rho-1<0,\medskip\\
\eta_{\min}=-\left(\dfrac{\tau_1\rho-1}{\rho}\right)^2-\tau_0^2-2+2\tau_0\tau_1:=\eta_4\\
\hspace{7mm}\mathrm{if} \rho^2-\tau_1\rho+1>0\ \mathrm{and}\ \rho^2+\tau_1\rho-1>0.
\end{array}
\ee
The condition for obtaining a PAAL for some points $(\xi_3,\xi_1)$ is hence, for $K=0$ orthotropic materials,
\be
\label{eq:condminetak0}
\eta_{\min}\in\{\eta_1,\eta_3,\eta_4\}<0.
\ee
The minimum $\eta_1$ is get in $\Omega$ at the point $(0,-1)$ and on the line $([-1,1],1)$, i.e. for any cross-ply combination and also for a laminate with all the layer orientations $\delta_j=0$ or $\delta_j=\dfrac{\pi}{2}$; $\eta_3$ is given by points $(\pm1,1)$, which is the case of a laminate with  the orientations of all the plies at the angle $\delta_j=0$ or $\delta_j=\dfrac{\pi}{2}$ and $\eta_4$ by the two symmetric points, on the boundary of $\Omega$, $\left(\pm\sqrt{\dfrac{\rho^2-\tau_1\rho+1}{2\rho^2}},\dfrac{1-\tau_1\rho}{\rho^2}\right)$,  that can be obtained by angle-ply sequences, but not only.

\begin{figure}
\centering
\includegraphics[width=.6\columnwidth]{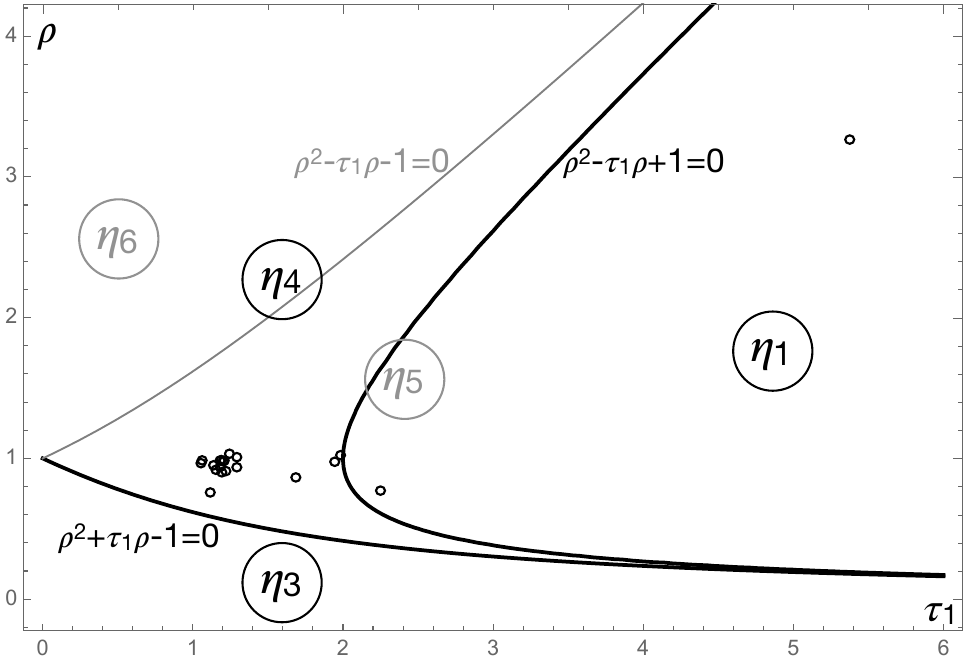}
\caption{The zones of the minimum of $\eta(\xi_3,\xi_1)$ for the case $K=0$, in black, and $K=1$, in grey. Each circle corresponds to a material of Tab. \ref{tab:1}.}
\label{fig:8}
\end{figure}

For $K=1$ orthotropic materials, it is
\be
\begin{array}{l}
\eta_{\min}=(\tau_0-\rho)(2\tau_1-\tau_0-\rho)-4:=\eta_5\\
\hspace{29mm} \mathrm{if}\ \ \rho^2-\tau_1\rho-1<0,\medskip\\
\eta_{\min}=-\left(\dfrac{\tau_1\rho+1}{\rho}\right)^2-\tau_0^2-2+2\tau_0\tau_1:=\eta_6\\
\hspace{38mm}\mathrm{if}\ \ \rho^2-\tau_1\rho-1>0.
\end{array}
\ee
The condition for obtaining a PAAL for some points $(\xi_3,\xi_1)$ is hence, for $K=1$ orthotropic materials,
\be
\eta_{\min}\in\{\eta_5,\eta_6\}<0.
\ee
The minimum $\eta_5$ is get at the points $(\pm1,1)$ i.e. by laminates with all the layer orientations $\delta_j=0$ or $\delta_j=\dfrac{\pi}{2}$; $\eta_6$ is given by the two symmetric points, on the boundary of $\Omega$, $\left(\pm\sqrt{\dfrac{\rho^2+\tau_1\rho+1}{2\rho^2}},\dfrac{1+\tau_1\rho}{\rho^2}\right)$,  also these lamination points can be obtained, but not only, by angle-ply sequences.
Just like  $\psi_{\min}$, also $\eta_{\min}$ depends on all the material parameters, but its position on $\Omega$ as well as the formula for its calculation depend only on $\tau_1$ and $\rho$.  The zones of the different $\eta_i,\ i=1,...,4,$ in the plane $(\tau_1,\rho)$ are shown in Fig. \ref{fig:8}, where, once more, the positions of the materials in Tab. \ref{tab:1} are also indicated by a small circle. Apart materials  7, a glass-epoxy ply, and  14, a composite boron-aluminium, all the layers belong to the zone where $\eta_{\min}=\eta_4$. For all of them, the minimum is get at a lamination point on the boundary of $\Omega$, that can always be obtained by an angle-ply sequence, but not exclusively.

\section{Materials suitable for the fabrication of auxetic laminates}
\begin{figure}
\centering
\includegraphics[width=.5\columnwidth]{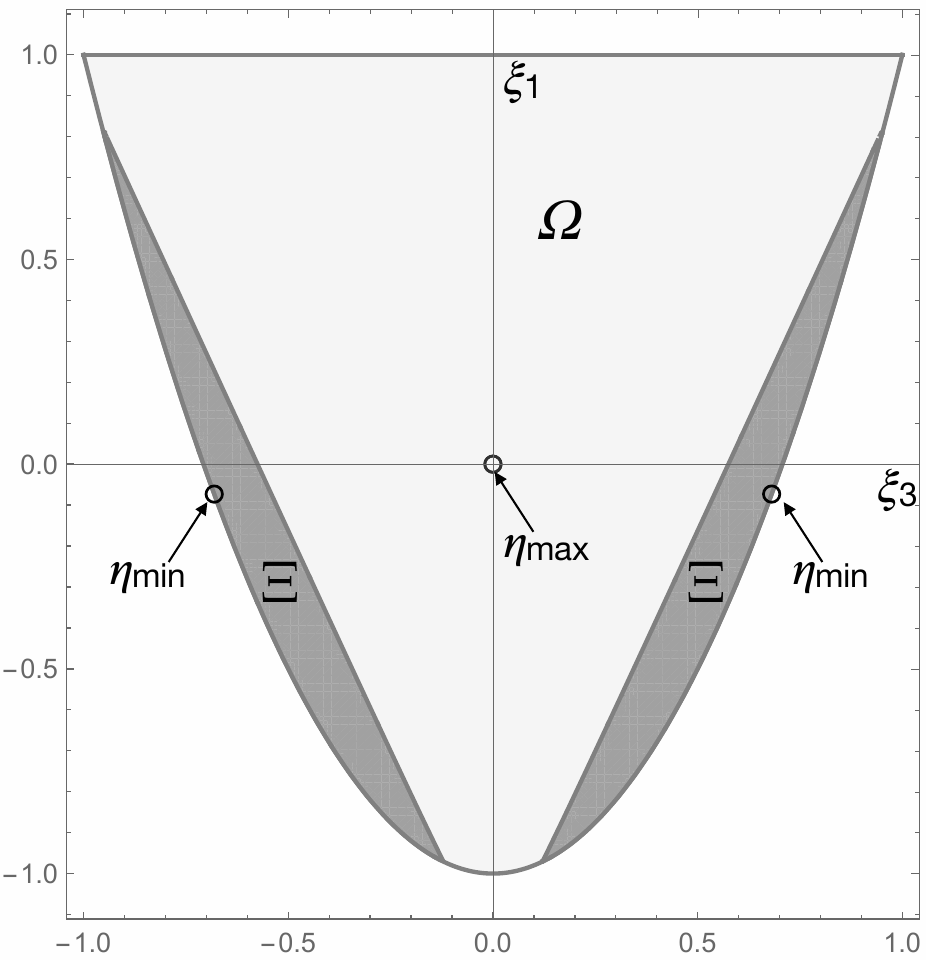}
\caption{Subset $\Xi\subset\Omega$ where  $\eta_{\min}<0$, material 2 in Tab. \ref{tab:1}.}
\label{fig:9}
\end{figure}

Looking at Figs. \ref{fig:3} and \ref{fig:6}, we can see that it is not guaranteed that functions $\psi(\xi_3,\xi_1)$ and $\eta(\xi_3,\xi_1)$ be negative on $\Omega$ or just on a subset of it. Physically, this means that it is not guaranteed that a TAAL or also a PAAL can always be realized, for any given anisotropic ply. For the examples in Figs. \ref{fig:3} and \ref{fig:6}, the only condition that is satisfied, on a subset  $\Xi\subset\Omega$, is that on $\eta_{\min}$ for material 2. So, for the cases at hand, only a PAAL can be realized, by laminates whose lamination point $(\xi_3,\xi_1)\in\Xi$ and made of material 2. The subset $\Xi$ for material 2 is shown in Fig. \ref{fig:9}.
Its existence, form and dimensions depend on the material properties $\tau_0,\tau_1$ and $\rho$.

The possibility of realizing a TAAL or a PAAL with a given layer is determined, in the space of the dimensionless polar moduli of the ply, by the conditions on the maximum or on the  minimum of $\psi(\xi_3,\xi_1)$ for the TAALs or of $\eta(\xi_3,\xi_1)$ for the PAALs, jointly to eq. (\ref{eq:thermo}). In the following, we rearrange such conditions for each one of the four cases treated above. 
So, for a given UD material with parameters $\tau_0,\tau_1,\rho$:
\begin{itemize}
\item it is possible to obtain a TAAL $\forall(\xi_3,\xi_1)\in\Omega\iff$
\be
\label{eq:domain1}
\left\{
\begin{array}{l}
\tau_0>\rho+2\tau_1,\medskip\\
\tau_1\left[\tau_0+(-1)^K\rho\right]-2>0.
\end{array}
\right.
\ee
The corresponding domain in the space $\{\tau_0,\tau_1,\rho\}$ is shown in Fig. \ref{fig:10}.
\begin{figure}
\centering
\includegraphics[width=.4\columnwidth]{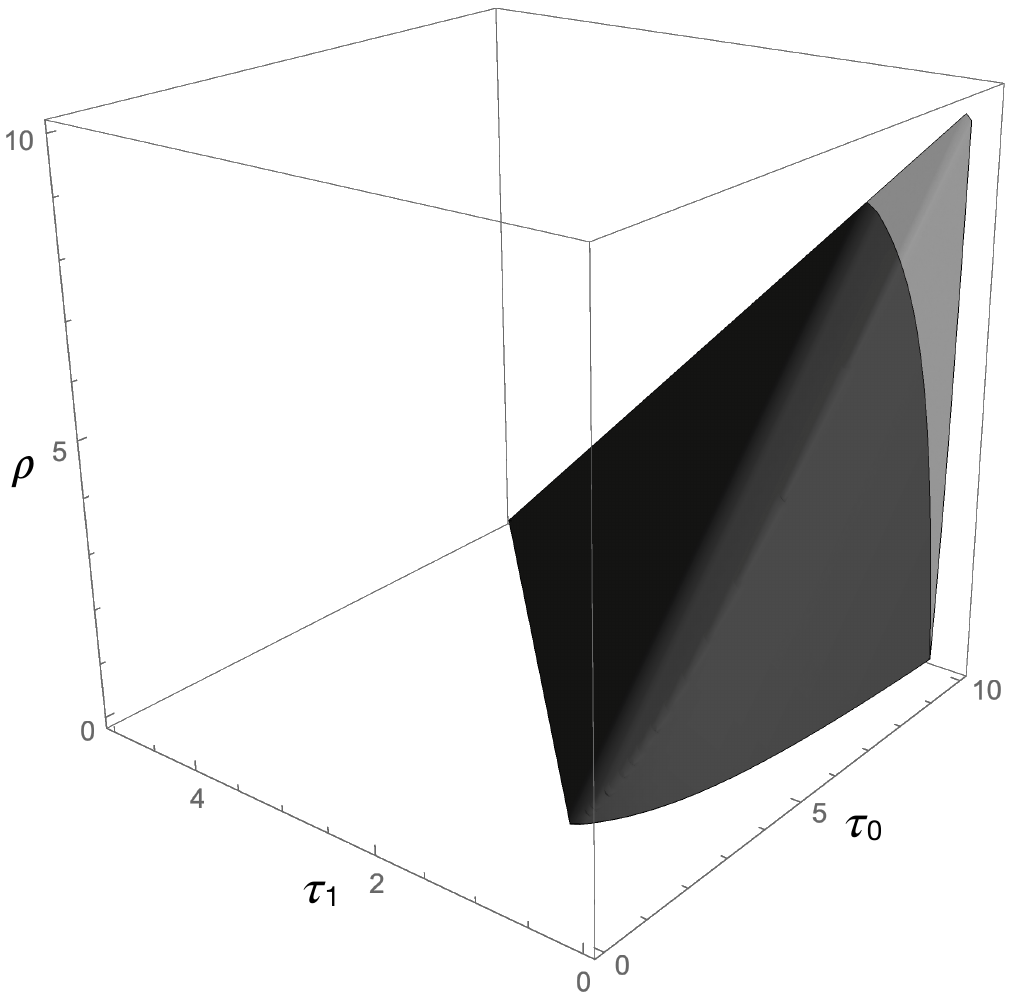}
\caption{Domain of  the space $\{\tau_0,\tau_1,\rho\}$ where  TAALs are possible $\forall(\xi_3,\xi_1)\in\Omega$. The exterior, light and transparent, domain corresponds to materials with $K=0$ and surrounds the, darker, part of materials with $K=1$.}
\label{fig:10}
\end{figure}
\item a TAAL can be obtained for some $(\xi_3,\xi_1)\in\Omega\iff$
\be
\label{eq:domain2}
\left\{
\begin{array}{l}
\tau_0-\rho>0,\medskip\\
\tau_1\left[\tau_0+(-1)^K\rho\right]-2>0,\medskip\\
\psi_{\min}\in\{\psi_1,\psi_2,\psi_3,\psi_4\}<0.
\end{array}
\right.
\ee
The domain in the space $\{\tau_0,\tau_1,\rho\}$ where TAALs are possible for some $(\xi_3,\xi_1)\in\Omega$ is shown in Fig. \ref{fig:11}.
\begin{figure}
\centering
\includegraphics[width=.4\columnwidth]{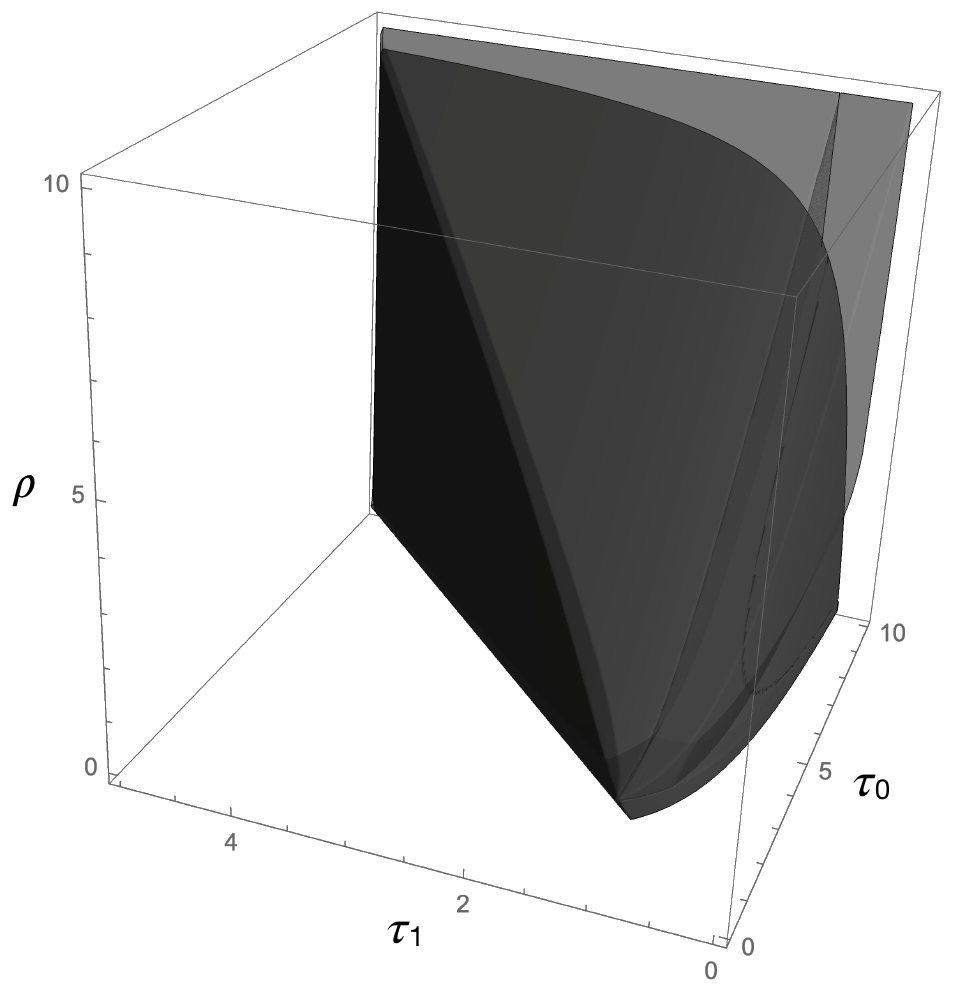}
\caption{Domain of  the space $\{\tau_0,\tau_1,\rho\}$ where  TAALs are possible for some $(\xi_3,\xi_1)\in\Omega$. The exterior, light and transparent, domain corresponds to materials with $K=0$ and surrounds the, darker, part of materials with $K=1$.}
\label{fig:11}
\end{figure}
\item a PAAL can be fabricated  $\forall(\xi_3,\xi_1)\in\Omega\iff$
\be
\label{eq:domain3}
\left\{
\begin{array}{l}
\tau_0-\rho>0,\medskip\\
\tau_1\left[\tau_0+(-1)^K\rho\right]-2>0,\medskip\\
\eta_{\max}\in\{\eta_1,\eta_2\}<0.
\end{array}
\right.
\ee
The domain in the space $\{\tau_0,\tau_1,\rho\}$ where PAALs are possible  $\forall(\xi_3,\xi_1)\in\Omega$ is shown in Fig. \ref{fig:12}. The part of the domain concerning materials with $K=0$  surrounds completely that for materials with $K=1$
\begin{figure}
\centering
\includegraphics[width=.4\columnwidth]{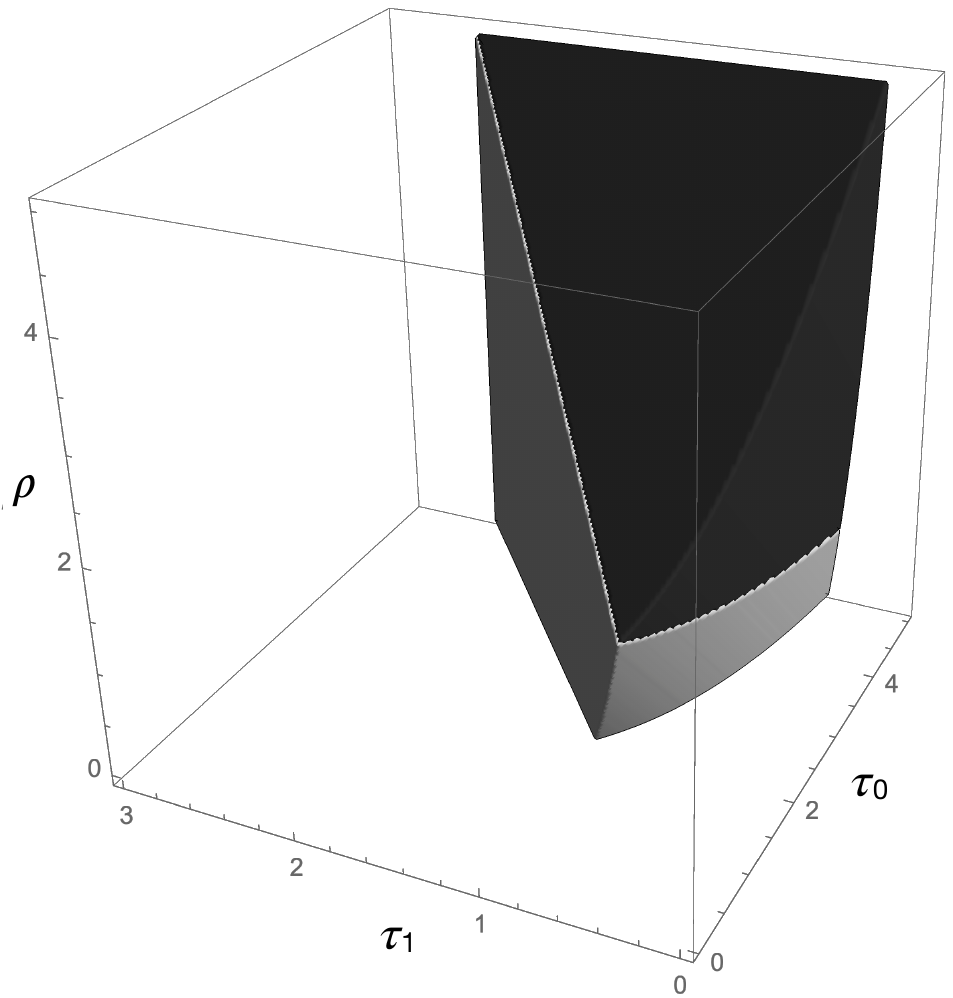}
\caption{Domain of  the space $\{\tau_0,\tau_1,\rho\}$ where  PAALs are possible  $\forall(\xi_3,\xi_1)\in\Omega$. The darker, upper part, $\rho>2\tau_1$,  corresponds to $\eta_{\max}=\eta_1$, the lighter, lower part, $\rho<2\tau_1$, to $\eta_{\max}=\eta_2$.}
\label{fig:12}
\end{figure}
\item  a PAAL can be fabricated  for some $(\xi_3,\xi_1)\in\Xi\subset\Omega\iff$
\be
\label{eq:domain4}
\left\{
\begin{array}{l}
\tau_0-\rho>0,\medskip\\
\tau_1\left[\tau_0+(-1)^K\rho\right]-2>0,\medskip\\
\eta_{\min}\in\{\eta_1,\eta_3,\eta_4,\eta_5,\eta_6\}<0.
\end{array}
\right.
\ee
The domain in the space $\{\tau_0,\tau_1,\rho\}$ where PAALs are possible  for some $(\xi_3,\xi_1)\in\Xi\subset\Omega$ is shown in Fig. \ref{fig:13}. 
\begin{figure}
\centering
\includegraphics[width=.4\columnwidth]{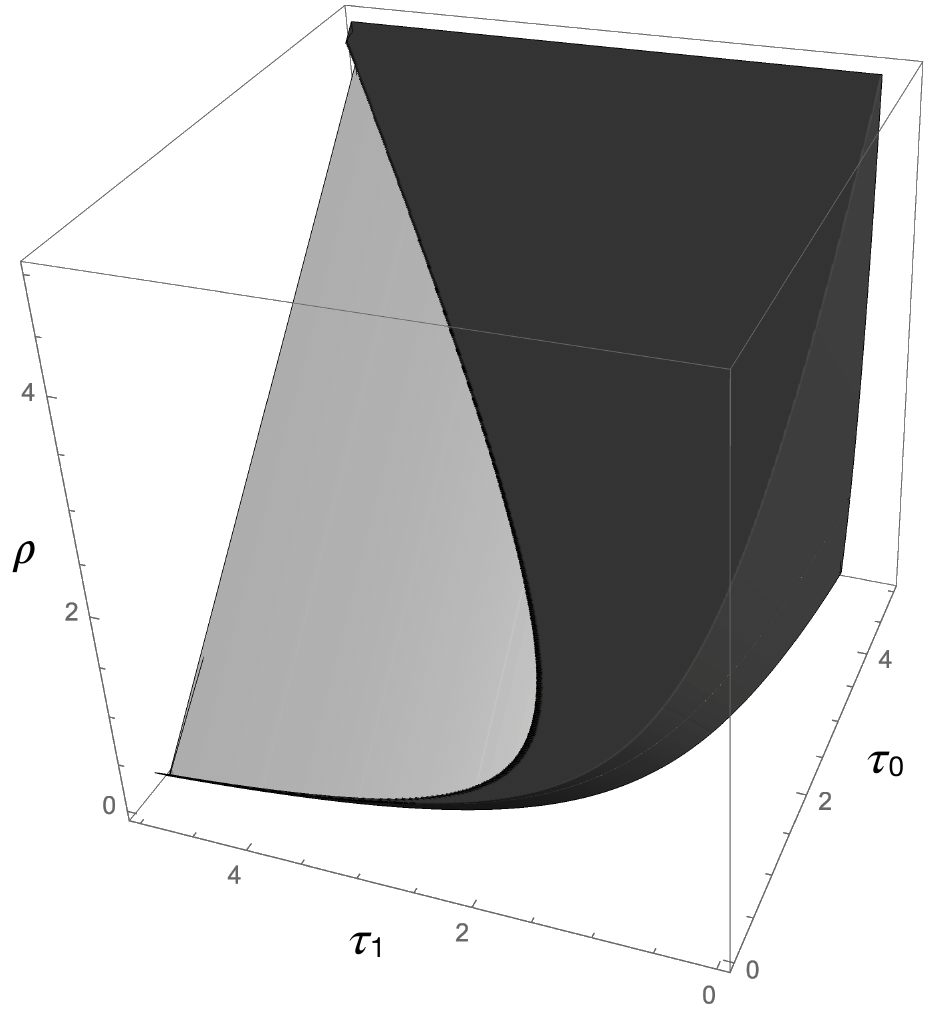}
\caption{Domain of  the space $\{\tau_0,\tau_1,\rho\}$ where  PAALs are possible  for some $(\xi_3,\xi_1)\in\Xi\subset\Omega$. The darker part corresponds to materials with $K=0$, the lighter one to $K=1$.}
\label{fig:13}
\end{figure}
\end{itemize}

The above results, represented in Figs. \ref{fig:10} to \ref{fig:13}, show that the four different types of auxetic laminates {\it can theoretically exist}: just like thermodynamics allows the existence of isotropic materials with a negative Poisson's ratio, in the same way it is physically possible to fabricate the four types of auxetic orthotropic laminates introduced above.

\section{Auxetic laminates made of non auxetic materials}
Although the results of the previous Section show that the four types of auxetic laminates can exist, a question still remains: a layer whose representative point in the space $\{\tau_0,\tau_1,\rho\}$ belongs to one of the domains in Figs. \ref{fig:10} to \ref{fig:13} can be fabricated using classical, i.e. non auxetic, isotropic components (fibres and matrix)? In other words, is it possible to obtain an auxetic laminate of one of the four types above using non auxetic materials? To give a  response to this question we need  to express  the polar moduli of $\Q$, i.e. $\tau_0, \tau_1, \rho$ and $K$, by the five parameters $E_f, E_m, \nu_f,\nu_m$ and $v_f$, respectively the Young's moduli of the fibres and of the matrix, the Poisson's ratio of the fibres and of the matrix, and the volume fraction of the fibres. Through a homogenization criterion, we first calculate the equivalent parameters of the layer (reference frame of the layer $\{x_1,x_2\}$, with $x_1$ aligned with the fibres): $E_1,E_2,G_{12}$ and $\nu_{12}$.

In this study, the matrix and the fibres are isotropic non auxetic materials and the classical rule of mixtures\cite{Jones,gay14} for UD layers is used to evaluate $E_1,E_2,G_{12}$ and $\nu_{12}$. In order to reduce the size of the problem, also for the technical moduli we introduce dimensionless parameters:
\be
E:=\frac{E_f}{E_m},\ \ \ \nu:=\frac{\nu_f}{\nu_m},\ \ \ G:=\frac{G_f}{G_m}
\ee
with (we suppose that the fibres are stiffer than the matrix)
\be
\label{eq:boundstechmoduli}
E>1,\ \ \ -\frac{1}{\nu_m}<\nu<\frac{1}{2\nu_m}.
\ee
In the above equations, $G_f$ and $G_m$ are, respectively, the shear moduli of the fibres and of the matrix, both assumed to be  isotropic:
\be
\label{eq:shearmoduli}
G_f=\frac{E_f}{2(1+\nu_f)},\ \ \ G_m=\frac{E_m}{2(1+\nu_m)}.
\ee
Through the rule of mixtures we get the following dimensionless constants for the layer:
\be
\label{eq:adimmoduli}
\bs
&\overline{E}_1:=\frac{E_1}{E_m}=1+v_f(E-1),\\
&\overline{E}_2:=\frac{E_2}{E_m}=\frac{E}{v_f+(1-v_f)E},\\
&\overline{\nu}_{12}:=\frac{\nu_{12}}{\nu_m}=1+v_f(\nu-1),\\
&\overline{G}_{12}:=\frac{G_{12}}{G_m}=\frac{G}{v_f+(1-v_f)G}=\\
&\hspace{7mm}=\frac{(1+\nu_m)E}{v_f(1+\nu\ \nu_m)+(1-v_f)(1+\nu_m)E}.
\end{split}
\ee

\begin{figure}
\centering
\includegraphics[width=.4\columnwidth]{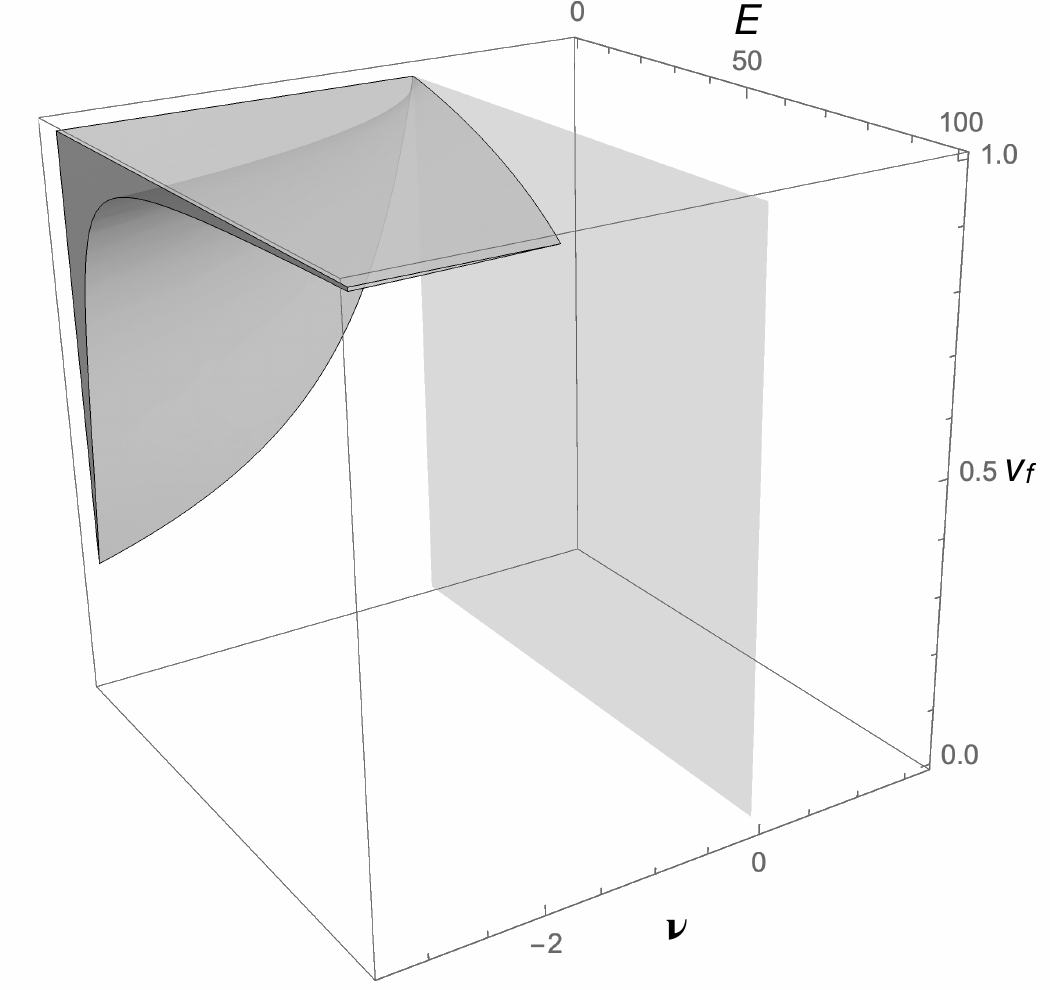}
\caption{Domain of  the space $\{E,\nu,v_f\}$ where  TAALs are possible $\forall(\xi_3,\xi_1)\in\Omega$.}
\label{fig:14}
\end{figure}

\begin{figure}
\centering
\includegraphics[width=.4\columnwidth]{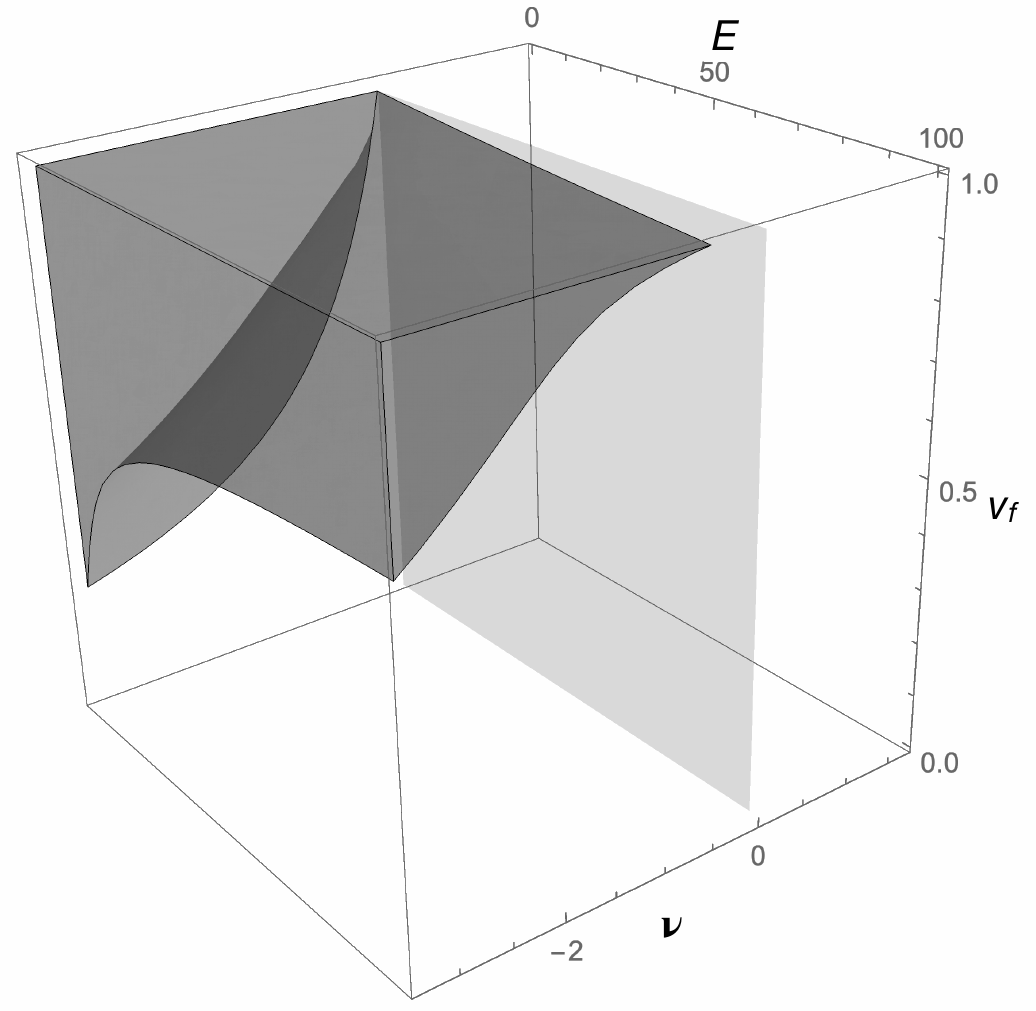}
\caption{Domain of  the space $\{E,\nu,v_f\}$ where  TAALs are possible for some $(\xi_3,\xi_1)\in\Omega$.}
\label{fig:15}
\end{figure}

\begin{figure}
\centering
\includegraphics[width=.4\columnwidth]{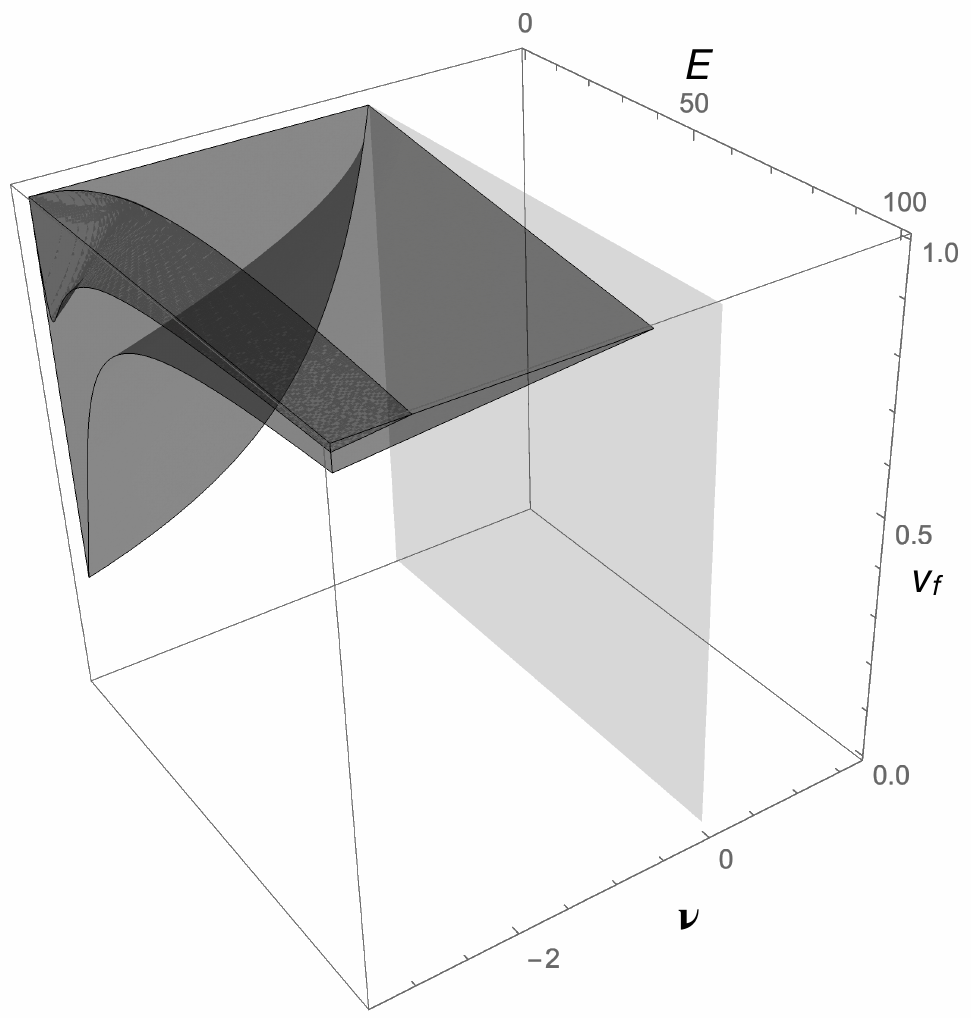}
\caption{Domain of  the space $\{E,\nu,v_f\}$ where  PAALs are possible $\forall(\xi_3,\xi_1)\in\Omega$.}
\label{fig:16}
\end{figure}

\begin{figure}
\centering
\includegraphics[width=.4\columnwidth]{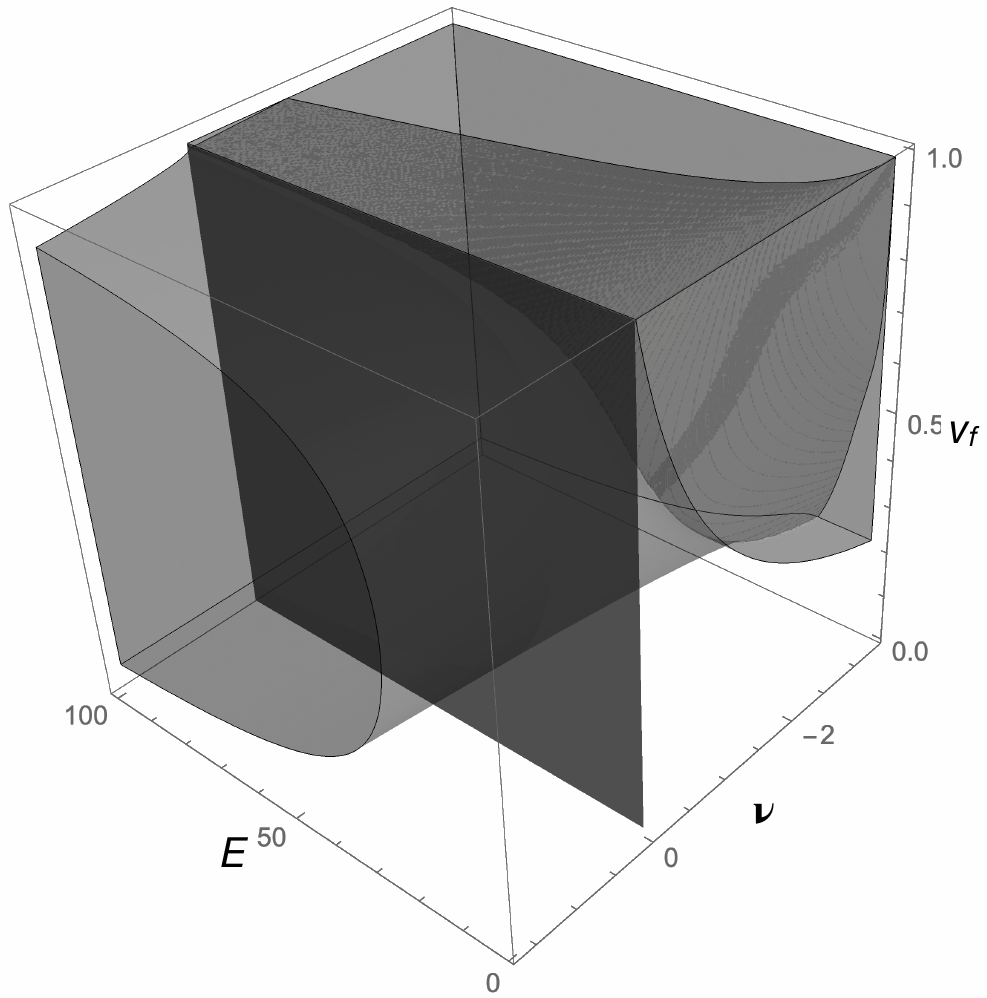}
\caption{Domain of  the space $\{E,\nu,v_f\}$ where  PAALs are possible for some $(\xi_3,\xi_1)\in\Xi\subset\Omega$.}
\label{fig:17}
\end{figure}

We can now calculate the components of  $\mathbb{S}$ and, by inversion, those of $\Q$, then used to calculate $T_0,T_1,(-1)^KR_0$ and $R_1$\cite{vannucci_libro} and finally $\tau_0,\tau_1,\rho$ and $K$. 
The mathematical details are given in the Appendix, where it is also shown that an UD material cannot have $K=1$; actually, all the UD plies in Tab. \ref{tab:1} have $K=0$.
The final expressions of $\tau_0,\tau_1,\rho$ as functions of $E,\nu,v_f$ for a given $\nu_m$ are:
\be
\label{eq:tau0ldm}
\bs
\tau_0&=
\frac{
\left(
\begin{array}{c}2 E \{(v_f-1) v_f+E^2 (v_f-1) v_f+\smallskip\\
E [2 v_f-2 v_f^2-1+(\nu_m+v_f (\nu-1 ) \nu_m)^2]\}
\end{array}
\right)}
{\left(
\begin{array}{c}(E-1)^2 (v_f-1) v_f [1+(E-1) v_f] \times\smallskip\\
\times[v_f+v_f
\nu  \nu_m-E (v_f-1) (1+\nu_m)]
\end{array}
\right)}+\\
&+\frac{2 E [1-\nu_m+v_f (v_f-1+\nu_m-\nu  \nu_m)]}{  v_f(1-v_f)(E-1)^2}+\frac{1+E^2}{(E-1)^2},
\end{split}
\ee
\be
\label{eq:tau1ldm}
\tau_1=\frac{
\left(
\begin{array}{c}(v_f-1) v_f+E^2 (v_f-1) v_f-\smallskip\\
-2 E \{1+\nu_m+v_f [v_f-1+(\nu-1) \nu_m]\}
\end{array}
\right)}
{ v_f(v_f-1)(E-1)^2 },
\ee 
\be
\label{eq:rholdm}
\rho=\frac{
\left(
\begin{array}{c}(E-1) [E (v_f-1)-v_f] [1+(E-1) v_f]+\smallskip\\
+(1+(E-1) v_f) [E^2 (v_f-1)+v_f \nu-\smallskip\\
 -E (1+v_f+(v_f-2) \nu )] \nu_m+\smallskip\\
+2 E [1+ (\nu-1 )v_f] (E-\nu ) \nu_m^2
\end{array}
\right)}
{\left(
\begin{array}{c}(E-1) [1+(E-1) v_f]\times\smallskip\\
\times [E (v_f-1) (1+\nu_m)-v_f (1+\nu  \nu_m)]
\end{array}
\right)}.
\ee
If the above expressions of $\tau_0,\tau_1,\rho$ are used in eqs. (\ref{eq:domain1}) to (\ref{eq:domain4}), the domain of existence of the four cases of auxetic orthotropic laminates can be traced in the space of the variables $E,\nu,v_f$, i.e. the corresponding of Figs. \ref{fig:10} to \ref{fig:13} in the space $\{E,\nu,v_f\}$ can be found. They are shown in Figs. \ref{fig:14} to \ref{fig:17} for a matrix having $\nu_m=0.3$ (the influence of this parameter is rather low and in any case does not vary much from one matrix to another).

In all these figures, the plane $\nu=0$ is shown: the parts of the domain belonging to the half-space $\nu<0$ corresponds to solutions that can be obtained only if one of the two constituent materials of the layer, the fibres or the matrix, is itself auxetic, that is, when one of the two constituents is not a classical, non auxetic elastic material. Looking at Figs. \ref{fig:14}-\ref{fig:16}, we see that the admissible domains are completely in the half-space $\nu<0$. This means that {\it it is impossible to fabricate auxetic orthotropic laminates that are completely auxetic or also partially auxetic for any possible orthotropic stacking sequence}. Only in Fig. \ref{fig:17} the feasible domain is for a part of it in the subspace $\nu>0$. Physically, this means that {\it only PAALs for some $(\xi_3,\xi_1)\in\Xi\subset\Omega$ can be fabricated using non auxetic layers}. It is worth to notice that this check could be done, theoretically, directly, adding the condition $\nu>0$ to the inequalities defining the domains for the four cases; unfortunately, the expressions are so complicate that this is impossible, in practice. Also, when tracing the domains in the space $\{E,\nu,v_f\}$ bounded by eqs. (\ref{eq:boundstechmoduli}), conditions (\ref{eq:thermo}) are automatically satisfied and can be discarded.

\section{Orthotropic PAALs}
Finally, two facts are crucial for obtaining auxetic orthotropic laminates: the material of the layers and the stacking sequence.
An UD layer is suitable for the realization of a PAAL only if condition (\ref{eq:condminetak0}) is satisfied. In fact, for a physical, existing, layer, the thermodynamic conditions (\ref{eq:thermo}) are necessarily satisfied, and because UD layers cannot have $K=1$, only condition (\ref{eq:condminetak0}) must be satisfied in order to obtain a PAAL for some $(\xi_3,\xi_1)$ in a subset $\Xi\subset\Omega$, i.e. for some stacking sequences. It is rather easy to check whether or not condition (\ref{eq:condminetak0}) is satisfied by a layer, and hence whether or not it is possible to realize with it auxetic laminates: it is sufficient to calculate its dimensionless polar  parameters $\tau_0,\tau_1$ and $\rho$ and then determine $\eta_{\min}$ and check if it is negative. In particular, for the materials in Tab. \ref{tab:1} it is always $\eta_{\min}=\eta_4$,  apart materials 7 and 14, that have $\eta_{\min}=\eta_1$. The value of $\eta_{\min}$ for all these materials is given in Tab. \ref{tab:2}.
\begin{table}
\centering\small\sf
\caption{Value of $\eta_{\min}$ for the materials of Tab. \ref{tab:1}.}
\begin{tabular}{rrrrrrrrrrr}
\toprule
Mat.&1&2&3&4&5\\
\midrule
$\eta_{\min}$&-0.86&-0.68&-0.52&0.49&-0.63\\
\bottomrule
Mat.&6&7&8&9&10\\
\midrule
$\eta_{\min}$&0.91&1.96&-0.87&-0.57&-0.42\\
\bottomrule
Mat.&11&12&13&14&15\\
\midrule
$\eta_{\min}$&0.92&-0.72&-0.62&2.52&-0.89\\
\bottomrule
Mat.&16&17&18&19&20\\
\midrule
$\eta_{\min}$&-0.54&-0.62&-0.59&-0.59&-0.38\\
\bottomrule
\end{tabular}
\label{tab:2}
\end{table}
Looking at the values of $\eta_{\min}$, we can see that it is not possible to obtain a PAAL with all the layers; namely, $\eta_{\min}>0$ for materials 4, 6, 7, 11 (glass-epoxy layers) and 14 (boron-aluminium). All of them are characterized by a low ratio ${E_1}/{E_2}$, whose value varies between 1.57 to 4.67. For all the other materials, having $\eta_{\min}<0$, this ratio varies from 9.85, materials 10 and 20, to 46.45, material 15. All these materials, apart material 1, pine wood, are composites obtained reinforcing a matrix with stiff fibres (carbon, kevlar, boron). It seems hence that auxeticity by anisotropy can be get only when anisotropy is sufficiently strong, practically, reinforcing a matrix with fibres that are much stiffer than it and with an adequate volume fraction. This is confirmed by Fig. \ref{fig:17}, where the part of the feasible domain with $\nu>0$ has high values of the ratio $E$.

\section{Naturally auxetic anisotropic plies}

\begin{figure}
\centering
\includegraphics[width=.5\columnwidth]{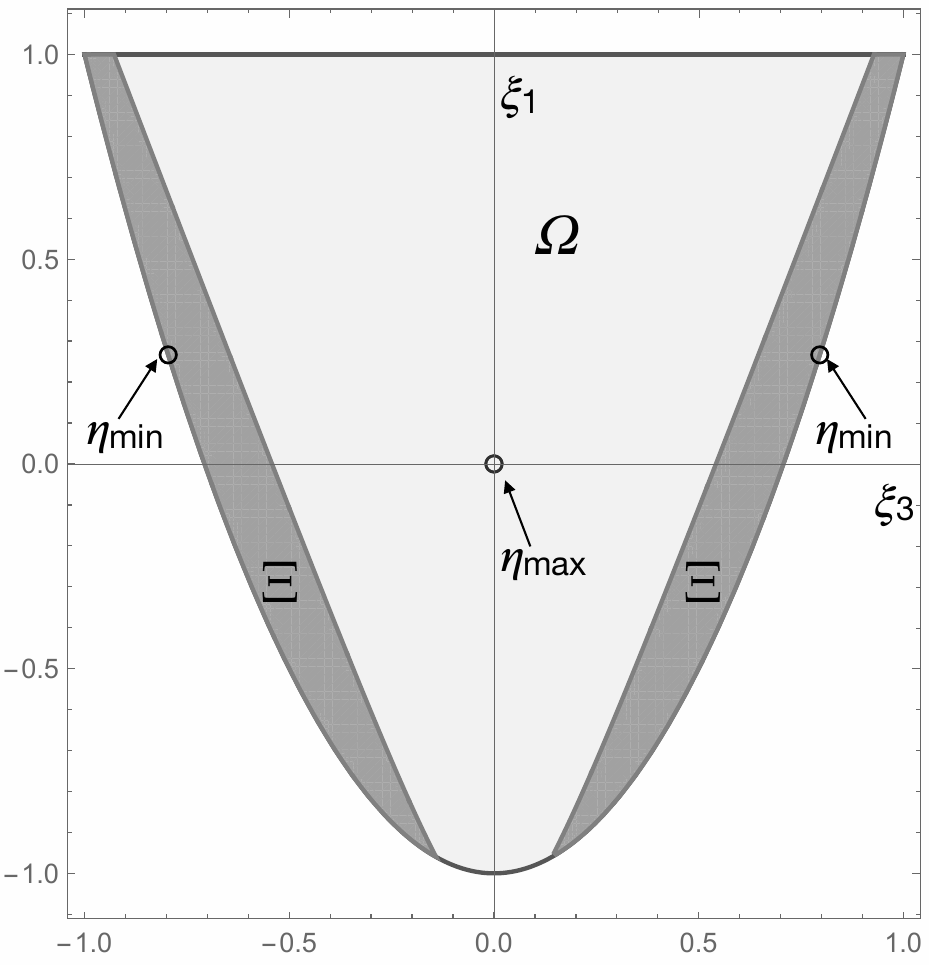}
\caption{Domain $\Xi$ for material 1 of Tab. \ref{tab:1}: pine wood.}
\label{fig:18}
\end{figure}

\begin{figure}
\centering
\includegraphics[width=\columnwidth]{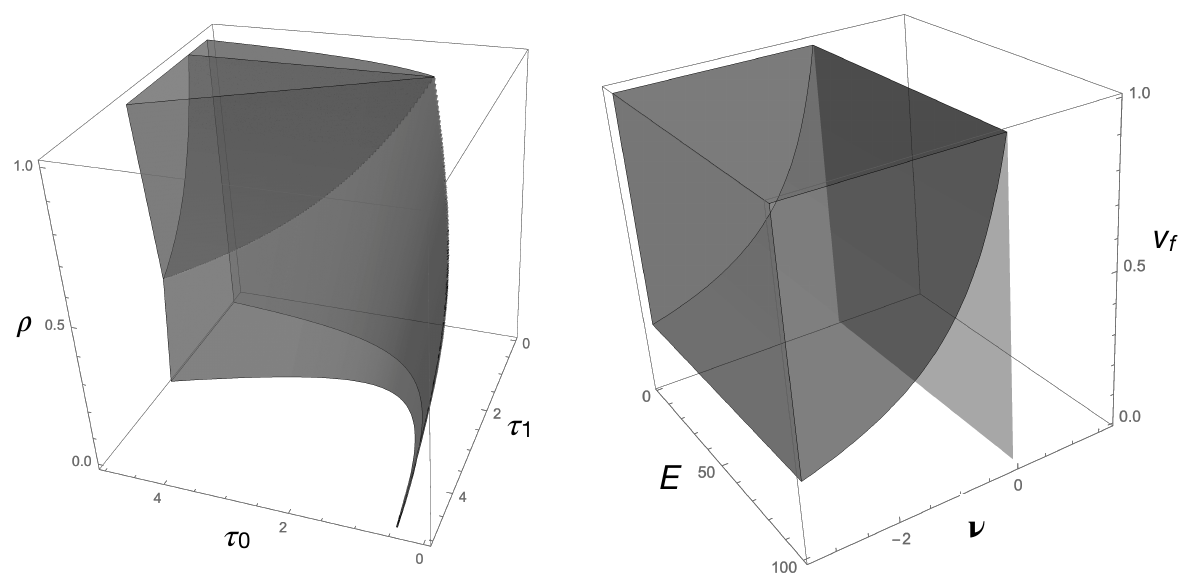}
\caption{The feasible domains of naturally anisotropic auxetic plies.}
\label{fig:21}
\end{figure}

The case of material 1, pine wood, is rather peculiar. Unlike all the other materials in Tab. \ref{tab:1}, it is a {\it naturally  auxetic anisotropic ply}, in the sense that, as already observed in the Introduction,  it is, by itself, already auxetic along some directions,  see Fig. \ref{fig:1}. This happens whenever the point $(1,1)\in\Xi$, see Fig. \ref{fig:18}.
The condition for an anisotropic ply  to be naturally auxetic can be get from equation (\ref{eq:auxUD}):  once divided by $R_1^2$ to obtain its dimensionless form, the condition is
\be
\begin{array}{c}
\mu:=\min\lambda(\theta)<0\Rightarrow\medskip\\
\mu=2(\tau_0\tau_1)-\tau_0^2+\rho^2-2|1-(-1)^K\tau_1\rho|<0,
\end{array}
\ee
which gives
\be
\begin{array}{l}
\mu_{\min}=(\tau_0+\rho)(2\tau_1-\tau_0+\rho)-4=\eta_3\medskip\\ \mathrm{if}\ \ K=0,\ \rho<\dfrac{1}{\tau_1},\medskip\\
\mu_{\min}=(\tau_0-\rho)(2\tau_1-\tau_0-\rho)=\eta_1\medskip\\ \mathrm{if}\ \ K=0,\ \rho>\dfrac{1}{\tau_1},\medskip\\
\mu_{\min}=(\tau_0-\rho)(2\tau_1-\tau_0-\rho)-4=\eta_5\medskip\\ \mathrm{if}\ \ K=1.
\end{array}
\ee
In the case of material 1, $
\rho=0.758<\dfrac{1}{\tau_1}=0.896\rightarrow\mu_{\min}=\eta_3=-0.559<0.
$
In Fig. \ref{fig:21} the admissible domains in the space $\{\tau_0,\tau_1,\rho\}$ and in the space $\{E,\nu,v_f\}$ are represented for the case $K=0$. It can be noticed that also in this case the part of the domain in the half space $\nu>0$ is null. This means that a UD cannot be naturally auxetic. The fact that material 1, pine wood, is naturally auxetic just means that its behavior cannot be reduced  to that of a UD ply whose characteristics are evaluated through the rule of mixtures.

\section{Some examples of PAALs}
We consider now some examples of PAALs made using the materials in Tab. \ref{tab:1}. In particular, as criterion for the choice, we have taken for each material the angle-ply corresponding to the point of the boundary of $\Xi$ where $\eta_{\min}$ is get. The angle, $\pm\delta$, of the plies is hence 
\be
\delta=\frac{1}{2}\arccos\xi_3=\frac{1}{4}\arccos\xi_1.
\ee
In Tab. \ref{tab:3} we list the cases considered, indicating for each of them $\eta_3,\eta_{\min}$, the coordinates $\xi_3^{\min},\xi_1^{\min}$ of $\eta_{\min}$, the orientation angle $\delta$ and  the maximum, $\nu_{12}^{\max}$, and minimum, $\nu_{12}^{\min}$,  of the Poisson's ratio.

The cases of materials 2, 5 and 15 show clearly that (partially) auxetic laminates can be obtained stacking UD layers that are not naturally auxetic, as a simple result of anisotropy.
\begin{table}
\centering\small\sf
\caption{Characteristics of some PAALs; the material numbers refer to Tab. \ref{tab:1}.}
\begin{tabular}{rrrrrrrr}
\toprule
Mat.&\hspace{-2mm}$\eta_3$&\hspace{-2mm}$\eta_{\min}$&\hspace{-2mm}$\xi_3^{\min}$&\hspace{-2mm}$\xi_1^{\min}$&\hspace{-2mm}$\delta$ ($^\circ$)&\hspace{-2mm}$\nud^{\max}$&\hspace{-2mm}$\nu_{12}^{\min}$\\
\midrule
1&\hspace{-2mm}-0.56&\hspace{-2mm}-0.86&\hspace{-2mm}$\pm$0.796&\hspace{-2mm}0.266&\hspace{-2mm}18.6&\hspace{-2mm}1.03&\hspace{-2mm}$-$0.40\\
2&\hspace{-2mm}0.29&\hspace{-2mm}-0.68&\hspace{-2mm}$\pm$0.681&\hspace{-2mm}$-$0.073&\hspace{-2mm}21.5&\hspace{-2mm}1.42&\hspace{-2mm}$-$0.32\\
5&\hspace{-2mm}0.69&\hspace{-2mm}-0.63&\hspace{-2mm}$\pm$0.645&\hspace{-2mm}$-$0.168&\hspace{-2mm}24.9&\hspace{-2mm}1.64&\hspace{-2mm}$-$0.34\\
15&\hspace{-2mm}0.08&\hspace{-2mm}-0.89&\hspace{-2mm}$\pm$0.700&\hspace{-2mm}$-$0.020&\hspace{-2mm}22.8&\hspace{-2mm}2.62&\hspace{-2mm}$-$0.93\\
\bottomrule
\end{tabular}
\label{tab:3}
\end{table}
\begin{figure}
\centering
\includegraphics[width=.5\columnwidth]{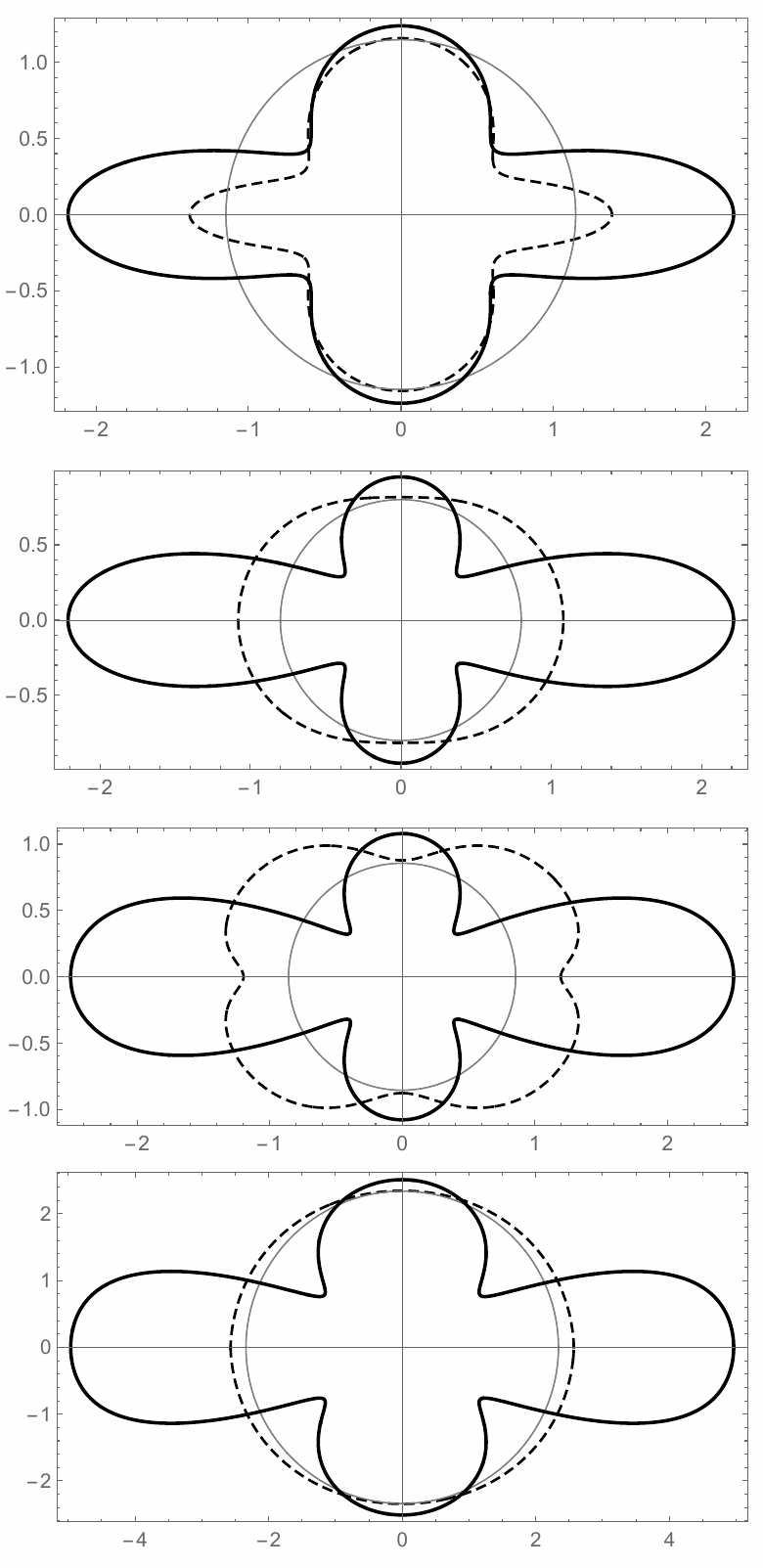}
\caption{Directional diagrams of $\nu_{12}$ for the four materials in Tab. \ref{tab:3}, respectively material 1, 2, 5 and 15 from top to bottom. The dashed line is $\nu_{12}$ of the ply. The thin  circle marks the zero value: inside it, $\nud$ is negative. }
\label{fig:19}
\end{figure}

\section{Conclusion}
The question of the possible realization of auxetic orthotropic laminates using layers made of non auxetic materials has been considered in this paper. The use of dimensionless polar parameters to define the material properties of the layer and of the lamination parameters for the description of the stack has allowed to define the admissible sets for auxetic laminates. Using the rule of mixture for the homogenization of the layer's properties allows  us to show that actually only partially auxetic laminates can be fabricated, for some admissible stacking sequences, using classical materials. This proves also that auxeticity can actually be produced simply by anisotropy. 

\section{Appendix}
In the Kelvin's notation\cite{kelvin,kelvin1}, the relations between the  polar parameters and the Cartesian components of $\Q$ for a UD ply are\cite{vannucci_libro}
\be
\begin{array}{l}
\label{eq:parpol}
T_0=\dfrac{1}{8}(\Q_{11} - 2 \Q_{12} + 2 \Q_{66} + \Q_{22}),\medskip\\
T_1=\dfrac{1}{8}(\Q_{11} +2 \Q_{12} + \Q_{22}),\medskip\\
(-1)^KR_0=\dfrac{1}{8}(\Q_{11} - 2 \Q_{12} -2 \Q_{66} + \Q_{22}),\medskip\\
R_1=\dfrac{1}{8}(\Q_{11} - \Q_{22}),\medskip\\
\end{array}
\ee
Because $R_0$ is intrinsically non negative, if $(-1)^KR_0<0$, then $K=1$, otherwise $K=0$.
The formulae for the polar parameters as functions of the engineering moduli $E_1,E_2,G_{12},\nu_{12}$ of the layer can be obtained expressing the $\Q_{ij}$s in terms of these last; for an UD layer
\be
\begin{array}{l}
\Q_{11}=\dfrac{E_1}{1-\nu_{12}\nu_{21}},\medskip\\
\Q_{12}=\dfrac{\nu_{12}E_2}{1-\nu_{12}\nu_{21}},\medskip\\
\Q_{22}=\dfrac{E_2}{1-\nu_{12}\nu_{21}},\medskip\\
\Q_{66}=2G_{12},
\end{array}
\ee
with $\nu_{21}$ given by the reciprocity relation $\nu_{21}=\nu_{12}\dfrac{E_2}{E_1}.$ Replacing these quantities into equation (\ref{eq:parpol}) gives
\be
\begin{array}{l}
T_0=\dfrac{1}{8} \left[\dfrac{E_1 (E_1+E_2-2 E_2 \nu_{12})}{E_1-E_2 \nu_{12}^2}+4 G_{12}\right],\medskip\\
T_1=\dfrac{1}{8}\dfrac{E_1 (E_1+E_2+2 E_2 \nu_{12})}{E_1-E_2 \nu_{12}^2},\medskip\\
(-1)^KR_0\hspace{-1mm}=\hspace{-1mm}\dfrac{1}{8}\hspace{-1mm} \left[\dfrac{E_1 (E_1+E_2-2 E_2 \nu_{12})}{E_1-E_2 \nu_{12}^2}-4 G_{12}\right],\medskip\\
R_1= \dfrac{1}{8} \dfrac{E_1 (E_1-E_2)}{E_1-E_2 \nu_{12}^2}.\medskip\\
\end{array}
\ee
The dimensionless parameters $\tau_0,\tau_1$ and $(-1)^K\rho$ can now be obtained:
\be
\begin{array}{l}
\tau_0=\dfrac{E_1 (E_1+E_2+4 G_{12})-2 E_1 E_2 \nu_{12}-4 E_2 G_{12} \nu_{12}^2}{E_1 (E_1-E_2)},\medskip\\
\tau_1=\dfrac{E_1+E_2+2 E_2 \nu_{12}}{E1-E2},\medskip\\
(-1)^K\rho\hspace{-1mm}=\hspace{-1mm}\dfrac{E_1 (E_1+E_2-4 G_{12})\hspace{-1mm}-\hspace{-1mm}2 E_1 E_2 \nu_{12}\hspace{-1mm}+\hspace{-1mm}4 E_2 G_{12} \nu_{12}^2}{E_1 (E_1-E_2)}.
\end{array}
\ee
If in these expressions we set
\be
\begin{array}{c}
E_1=\overline{E}_1E_m,\ \ E_2=\overline{E}_2E_m,\medskip\\ \nu_{12}=\overline{\nu}_{12}\nu_m,\ \ G_{12}=\overline{G}_{12}G_m,
\end{array}
\ee
with $\overline{E}_1,\overline{E}_2,\overline{\nu}_{12},\overline{G}_{12}$ and $G_m$ given by equations (\ref{eq:shearmoduli}) and (\ref{eq:adimmoduli}), then equations (\ref{eq:tau0ldm}) to (\ref{eq:rholdm}) are get, giving the polar dimensionless parameters $\tau_0,\tau_1$ and $\rho$ as functions of the dimensionless constants $E,\nu,v_f$ and of the Poisson's ratio of the matrix, $\nu_m$.

\begin{figure}
\centering
\includegraphics[width=.4\columnwidth]{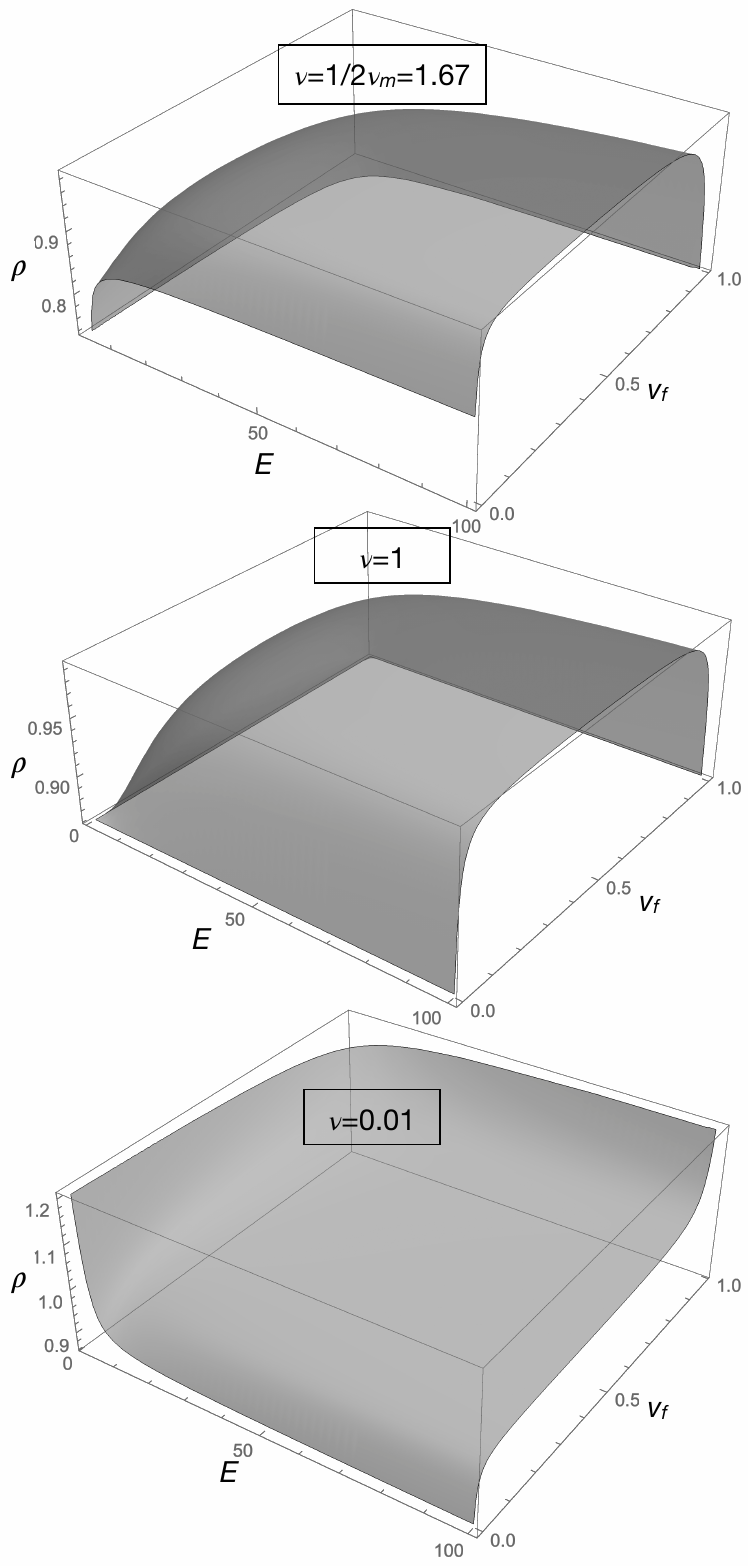}
\caption{The parameter $(-1)^K\rho$ for three different ratios $\nu$ and for $\nu_m=0.3$.}
\label{fig:20}
\end{figure}

In Fig. \ref{fig:20} the parameter $(-1)^K\rho$ is shown for three different ratios $\nu$ and for a matrix with $\nu_m=0.3$. It is apparent that in all the cases $(-1)^K\rho>0\ \ \forall(E,v_f)$. Because $\rho$ is intrinsically a non negative quantity, $\rho>0\Rightarrow K=0$. Physically, this means that a UD ply cannot have $K=1$.

\bibliographystyle{SageV}
\bibliography{Bibliotot.bib}

\begin{thebibliography}{10}
\providecommand{\url}[1]{\texttt{#1}}
\providecommand{\urlprefix}{URL }
\expandafter\ifx\csname urlstyle\endcsname\relax
  \providecommand{\doi}[1]{DOI:\discretionary{}{}{}#1}\else
  \providecommand{\doi}{DOI:\discretionary{}{}{}\begingroup
  \urlstyle{rm}\Url}\fi
\providecommand{\eprint}[2][]{\url{#2}}

\bibitem{cho2019}
Cho H, Seo D and Kim DN.
\newblock Mechanics of auxetic materials.
\newblock In Schmauder S and Chen CS (eds.) \emph{Handbook of mechanics of
  materials}. Singapore: Springer, 2019.
\newblock pp. 733--757.

\bibitem{Almgren85}
Almgren RF.
\newblock {An isotropic three-dimensional structure with Poisson's ratio= -1}.
\newblock \emph{Journal of Elasticity} 1985; 15: 427--430.

\bibitem{Evans91}
Evans KE.
\newblock Auxetic polymers: a new range of materials.
\newblock \emph{Endeavour - New Series} 1991; 15(4): 170--174.

\bibitem{Evans_Nature91}
Evans KE, Nkansah MA, Hutchinson IJ et~al.
\newblock Molecular network design.
\newblock \emph{Nature} 1991; 353: 124.

\bibitem{Lakes1991}
Lakes R.
\newblock {Deformation mechanisms in negative Poisson's ratio materials:
  structural aspects}.
\newblock \emph{Journal of Materials Sciences} 1991; 26: 2287--2292.

\bibitem{Lakes1993}
Lakes R.
\newblock {Advances in negative Poisson's ratio materials}.
\newblock \emph{Advanced Materials} 1993; 5(4): 293--296.

\bibitem{Lakes2002}
Lakes R and Witt R.
\newblock {Making and characterizing negative Poisson's ratio materials}.
\newblock \emph{International Journal of Mechanical Engineering Education}
  2002; 30(1): 50--58.

\bibitem{Lakes2017}
Lakes R.
\newblock {Negative Poisson's-ratio materials: Auxetic solids}.
\newblock \emph{Annaul Review of Materials Research} 2017; 47: 63--81.

\bibitem{Milton92}
Milton GW.
\newblock {Composite materials with Poisson's ratios close to -1}.
\newblock \emph{Journal of the Mechanics and Physics of Solids} 1992; 40(5):
  1105--1137.

\bibitem{Prawoto12}
Prawoto Y.
\newblock {Seeing auxetic materials from the mechanics point of view: A
  structural review on the negative Poisson's ratio}.
\newblock \emph{Computational Materials Science} 2012; 58: 140--153.

\bibitem{Shukla22}
Shukla S and Behera BK.
\newblock {Auxetic fibrous structures and their composites: A review}.
\newblock \emph{Composite Structures} 2022; 290(115530).

\bibitem{Lekhnitskii}
Lekhnitskii SG.
\newblock \emph{Theory of elasticity of an anisotropic elastic body}.
\newblock San Francisco, CA: English translation (1963) by {P. Fern},
  Holden-Day, 1950.

\bibitem{Veloso23}
Veloso C, Mota C, Cunha F et~al.
\newblock A comprehensiv review on in-plane and through-the-thickness
  auxeticity in composite laminates for structural applications.
\newblock \emph{Journal of Composite Materials} 2023; 57(26): 4215--4223.

\bibitem{Herakovich84}
Herakovich CT.
\newblock {Composite laminates with negative through-the-thickness Poisson's
  ratios}.
\newblock \emph{Journal of Composite Materials} 1984; 18(5): 447--455.

\bibitem{Miki89}
Miki M and Murotsu Y.
\newblock {The peculiar behavior of the Poisson's ratio of laminated fibrous
  composites}.
\newblock \emph{JSME International Journal} 1989; 32(1): 67--72.

\bibitem{Clarke94}
Clarke JJ, Duckett RA, Hine PJ et~al.
\newblock {Negative Poisson's ratios in angle-ply laminates: theory and
  experiment}.
\newblock \emph{Composites} 1994; 25(9): 863--868.

\bibitem{Hine97}
Hine PJ, Duckett RA and Ward IM.
\newblock {Negative Poisson's ratios in angles-ply laminates}.
\newblock \emph{Journal of Materials Science Letters} 1997; 16: 541--544.

\bibitem{Zhang98}
Zhang R, Yeh HL and Yeh HY.
\newblock {A preliminary study of negative Poisson's ratio of laminated fiber
  reinforced composites}.
\newblock \emph{Journal of Reinforced Plastics and Composites} 1998; 17(18):
  1651--1664.

\bibitem{Zhang99}
Zhang R, Yeh HL and Yeh HY.
\newblock {A discussion of negative Poisson's ratio design for composites}.
\newblock \emph{Journal of Reinforced Plastics and Composites} 1999; 18(17):
  1546--1556.

\bibitem{alderson2005}
Alderson KL, Simkins VR, Coenen VL et~al.
\newblock How to make auxetic fibre reinforced composites.
\newblock \emph{Physica Status Solidi} 2005; 242(3): 509--518.

\bibitem{Peel07}
Peel LD.
\newblock {Exploration of high and negative Poisson's ratio elastomer-matrix
  laminates}.
\newblock \emph{Physica Status Solidi} 2007; 244(3): 988--1003.

\bibitem{Shokrieh11}
Shokrieh MM and Assadi A.
\newblock {Determination of maximum negative Poisson's ratio for laminated
  fiber composites}.
\newblock \emph{Physica Status Solidi} 2011; 248(5): 1237--1241.

\bibitem{Verchery79}
Verchery G.
\newblock Les invariants des tenseurs d'ordre 4 du type de
  l'{\'e}lasticit{\'e}.
\newblock In \emph{{Proc. of Colloque Euromech 115 (Villard-de-Lans, 1979):
  Comportement m{\'e}canique des mat{\'e}riaux anisotropes}}. Paris: Editions
  du CNRS, 1982.
\newblock pp. 93--104.

\bibitem{Vannucci05}
Vannucci P.
\newblock Plane anisotropy by the polar method.
\newblock \emph{Meccanica} 2005; 40: 437--454.

\bibitem{vannucci_libro}
Vannucci P.
\newblock \emph{Anisotropic elasticity}.
\newblock Berlin, Germany: Springer, 2018.

\bibitem{tsai1968}
Tsai SW and Pagano NJ.
\newblock Invariant properties of composite materials.
\newblock In Tsai SW, Halpin JC and Pagano NJ (eds.) \emph{Composite Materials
  Workshop}. Stamford, CT: Technomic, 1968.

\bibitem{TsaiHahn}
Tsai SW and Hahn T.
\newblock \emph{Introduction to composite materials}.
\newblock Stamford, CT: Technomic, 1980.

\bibitem{Jones}
Jones RM.
\newblock \emph{Mechanics of composite materials. Second Edition}.
\newblock Philadelphia, PA: Taylor \& Francis, 1999.

\bibitem{Ting}
Ting TCT.
\newblock \emph{Anisotropic elasticity}.
\newblock Oxford, UK: Oxford University Press, 1996.

\bibitem{kelvin}
{Thomson - Lord Kelvin} W.
\newblock Elements of a mathematical theory of elasticity.
\newblock \emph{Philosophical Transations of the Royal Society} 1856; 146:
  481--498.

\bibitem{kelvin1}
{Thomson - Lord Kelvin} W.
\newblock Mathematical theory of elasticity.
\newblock \emph{Encyclopedia Britannica} 1878; 7: 819--825.

\bibitem{vannucci01joe}
Vannucci P.
\newblock On bending-tension coupling of laminates.
\newblock \emph{Journal of Elasticity} 2001; 64: 13--28.

\bibitem{vannucci23a}
Vannucci P.
\newblock On the mechanical and mathematical properties of the stiffness and
  compliance coupling tensors of composite anisotropic laminates.
\newblock \emph{Journal of Composite Materials} 2023; 57(26): 4197--4214.

\bibitem{vannucci23b}
Vannucci P.
\newblock On the thermoelastic coupling of anisotropic laminates.
\newblock \emph{Archive of Applied Mechanics} 2024; to appear.

\bibitem{vannucci01ijss}
Vannucci P and Verchery G.
\newblock Stiffness design of laminates using the polar method.
\newblock \emph{International Journal of Solids and Structures} 2001; 38:
  9281--9294.

\bibitem{vannucci01cst}
Vannucci P and Verchery G.
\newblock A special class of uncoupled and quasi-homogeneous laminates.
\newblock \emph{Composites Science and Technology} 2001; 61: 1465--1473.

\bibitem{vannucci13}
Vannucci P.
\newblock A note on the elastic and geometric bounds for composite laminates.
\newblock \emph{Journal of Elasticity} 2013; 112: 199--215.

\bibitem{gay14}
Gay D.
\newblock \emph{{Composite Materials Design and Applications - Third Edition}}.
\newblock Boca Raton, FL: CRC Press, 2014.

\bibitem{daniel94}
Daniel IM and Ishai O.
\newblock \emph{Engineering mechanics of composite materials}.
\newblock Oxford, UK: Oxford University Press, 1994.

\bibitem{MILHDBK}
AAVV.
\newblock {MIL-HDBK - Composite Materials Handbook - Volume 2}.
\newblock Technical report, US Department of Defense, 2002.

\bibitem{Gurdal99}
G{\"u}rdal Z, Haftka RT and Hajela P.
\newblock \emph{Design and optimization of laminated composite materials}.
\newblock New York, NY: J. Wiley \& Sons, 1999.

\bibitem{Love}
Love AEH.
\newblock \emph{A treatise on the mathematical theory of elasticity}.
\newblock New York, NY: Dover, 1944.

\bibitem{vannucci24}
Vannucci P.
\newblock {Complete set of bounds for the technical moduli in 3D anisotropic
  elasticity}.
\newblock \emph{Journal of Elasticity} 2024; to appear.

\bibitem{Miki82}
Miki M.
\newblock Material design of composite laminates with required in-plane elastic
  properties.
\newblock In \emph{Proc. of ICCM 4 - Fourth International Conference on
  Composite Materials}. Tokio, Japan, 1982.
\newblock pp. 1725--1731.

\bibitem{Miki83}
Miki M.
\newblock A graphical method for designing fibrous laminated composites with
  required in-plane stiffness.
\newblock \emph{Transactions of the Japanese Society of Composite Materials}
  1983; 9: 51--55.

\bibitem{Miki85}
Miki M.
\newblock Design of laminated fibrous composite plates with required flexural
  stiffness.
\newblock In Vinson JR and Taya M (eds.) \emph{Recent advances in composites in
  the USA and Japan - ASTM STP 864}. Philadelphia, PA: American Society for
  Testing and Materials, pp. 387--400.

\end{thebibliography}

\end{document}